\documentclass[conference]{IEEEtran}
\IEEEoverridecommandlockouts

\usepackage{cite}
\usepackage{amsmath,amssymb,amsfonts}
\usepackage{graphicx}
\usepackage{textcomp}
\usepackage{xcolor}
\usepackage{caption}
\usepackage{verbatim}
\usepackage{subfig}
\usepackage{multirow}
\usepackage{setspace}

\usepackage{url}
\usepackage{breakurl}
\usepackage[hidelinks]{hyperref}
\usepackage{bbm}
\usepackage{bm}

\usepackage{algorithm}
\usepackage{algorithmic}

\usepackage{graphicx}
\usepackage{textcomp}

\usepackage{theorem}
\theoremstyle{plain}
\newtheorem{theorem}{Theorem}
\newtheorem{definition}{Definition}
\newtheorem{lemma}{Lemma}

\usepackage{fancyhdr}
\fancypagestyle{firststyle}
{
	\fancyhf{}
	
	\fancyhead[C]{\emph{This is a full version of our paper that appears in the 36th IEEE International Conference on Data Engineering (ICDE 2020)}}
	\fancyfoot[C]{\thepage}
}

\begin{document}

\title{Providing Input-Discriminative Protection \\ for Local Differential Privacy \thanks{This work was partly supported by NSF grants CNS-1731164 and CNS-1618932, Air Force Office of Scientific Research (AFOSR) DDDAS program under grant FA9550-12-1-0240, JSPS KAKENHI grants with number 17H06099, 18H04093, 19K20269, and Microsoft Research Asia.}}

\author{\IEEEauthorblockN{Xiaolan Gu}
\IEEEauthorblockA{\textit{Department of ECE} \\
\textit{University of Arizona}\\
Tucson, AZ, USA \\
xiaolang@email.arizona.edu}
\and
\IEEEauthorblockN{Ming Li}
\IEEEauthorblockA{\textit{Department of ECE} \\
\textit{University of Arizona}\\
Tucson, AZ, USA \\
lim@email.arizona.edu}
\and
\IEEEauthorblockN{Li Xiong}
\IEEEauthorblockA{\textit{Department of Computer Science} \\
\textit{Emory University}\\
Atlanta, GA, USA \\
lxiong@emory.edu}
\and
\IEEEauthorblockN{Yang Cao}
\IEEEauthorblockA{\textit{Department of Social Informatics} \\
\textit{Kyoto University}\\
Kyoto, Japan \\
yang@i.kyoto-u.ac.jp}
}

\maketitle

\pagestyle{plain}
\thispagestyle{firststyle}

\begin{abstract}
Local Differential Privacy (LDP) provides provable privacy protection for data collection without the assumption of the trusted data server. In the real-world scenario, different data have different privacy requirements due to the distinct sensitivity levels. However, LDP provides the same protection for all data. In this paper, we tackle the challenge of providing input-discriminative protection to reflect the distinct privacy requirements of different inputs. We first present the Input-Discriminative LDP (ID-LDP) privacy notion and focus on a specific version termed MinID-LDP, which is shown to be a fine-grained version of LDP. Then, we focus on the application of frequency estimation and develop the IDUE mechanism based on Unary Encoding for single-item input and the extended mechanism IDUE-PS (with Padding-and-Sampling protocol) for item-set input. The results on both synthetic and real-world datasets validate the correctness of our theoretical analysis and show that the proposed mechanisms satisfying MinID-LDP have better utility than the state-of-the-art mechanisms satisfying LDP due to the input-discriminative protection.
\end{abstract}


\section{Introduction}

Differential Privacy (DP) \cite{dwork2006differential,dwork2006calibrating} has become the \emph{de facto} standard for private data release. It provides provable privacy protection, which is independent of the adversary's background knowledge and computational power \cite{chen2016private}.  In recent years,  Local Differential Privacy (LDP) has been proposed for preserving privacy at the data collection stage, in contrast to DP in the centralized setting which protects data after it is collected and stored by a server. In the local setting, the server is assumed to be untrusted, and each user randomly perturbs her raw data independently using a privacy-preserving mechanism that satisfies LDP. Then, the server collects these perturbed data from all users to perform data analytics or answer queries from users or third parties. Thus the local setting has been widely adopted in practice. For example, RAPPOR \cite{erlingsson2014rappor} proposed by Google has been employed in Chrome to collect web browsing behavior with LDP guarantees; Apple is also using LDP-based mechanism to identify popular emojis and popular health data types in Safari \cite{apple2017learning}.

Under the notion of LDP, given any output of a mechanism, the adversary cannot distinguish any pair of inputs with high confidence (controlled by a privacy budget $\epsilon$). Due to the uniform privacy budget, existing LDP mechanisms and applications \cite{erlingsson2014rappor,wang2017locally,wang2018locally,ye2019privkv} would perturb the data in the same way (or add the noise with the same amount) for any inputs. However, in many practical scenarios, different inputs have different degrees of sensitivity (i.e., users' desired privacy level or privacy expectation on the raw data) thus require distinct levels of privacy. For example, in website-click records or medical records, some website pages or medical diseases (e.g., HIV and  cancer) are much more sensitive than others, thus need stronger privacy guarantees; on the other hand, some records are much less sensitive, such as commonly visited pages by many people (e.g., Facebook and Amazon), or some  very common symptoms in clinic such as anemia and headache. Existing notions do not deal with this scenario. For example,  personalized local differential privacy (PLDP) \cite{wang2015personalized,chen2016private} only provides user-level discrimination, and geo-indistinguishability \cite{andres2013geo} only provides distance based discrimination for a pair of locations. 

Motivated by such considerations, we consider the categorical data and assume the universe of inputs have multiple levels of privacy, represented by privacy budgets with different values. Note that a smaller budget indicates higher privacy requirement thus needs more protection. In practice, classifying items by privacy levels can be implemented according to some categories with semantic meanings. For example, serious diseases (e.g., various cancers or HIV) can be classified in the highest privacy level, while moderate diseases (e.g., asthma or hypertension) and common symptoms can be classified in the medium and lowest privacy levels respectively. Since each possible input $x$ in domain $\mathcal{D}$ has its privacy budget $\epsilon_x$ (inputs with the same privacy level have the same budget), the privacy budget of standard LDP should be  $\epsilon=\min_{x\in\mathcal{D}}\{\epsilon_x\}$ to satisfy the required privacy for all inputs. Thus, LDP would provide excessive protection for some inputs that do not need such strong privacy, which is unnecessary and will lead to an inferior privacy-utility tradeoff. 

In this paper, we aim at providing input-discriminative privacy with distinct protection for each input and high utility on frequency estimation. We first study how to formalize a privacy notion in the local setting that provides discriminative privacy protection for different inputs. We propose a notion called Input-Discriminative LDP (ID-LDP) by converting the differentiated protection for inputs into different indistinguishability level for pairs of inputs. Theoretically, the indistinguishability of a pair of inputs $x,x^{\prime}$  can be any function of their budgets $\epsilon_x$ and $\epsilon_{x^{\prime}}$. In this paper, we focus on one instantiation termed MinID-LDP with the minimum function. It relaxes LDP on the inputs that do not need too strong privacy protection, and we will show that  the relaxation is at most twice of the minimum privacy budget of standard LDP (in Lemma \ref{lem:relationship}). In summary, MinID-LDP can provide fine-grained protection where each input is protected with required indistinguishability, while LDP would overprotect the inputs that have less sensitivity. 

Under our MinID-LDP notion, users need to perturb different inputs with different parameters related to the distinct privacy budgets, which makes the problem complicated since the perturbation parameters of a specific input may also depend on other inputs' privacy budgets to achieve indistinguishability between any two possible inputs.  To find the optimal mechanism for a real-world query function, a potential solution is to formulate an optimization problem with the goal of maximizing query utility given privacy as constraints. However, the objective function of minimizing the Mean Squared Error (MSE) of the unbiased estimator is dependent on the unknown true frequencies thus cannot be directly evaluated. Also, the computation complexity is high because MinID-LDP considers multiple different privacy budgets, which leads to large numbers of variables (perturbation parameters need to be solved) and privacy constraints (which should be satisfied for any  inputs $x,x^{\prime}$ and output $y$).


In this paper, we design two efficient and near-optimal mechanisms satisfying ID-LDP for frequency estimation on single-item and item-set data respectively. First, we propose Input-Discriminative Unary Encoding (IDUE) mechanism for \emph{single-item input}. The objective function in optimization problem of assigning the perturbation probabilities in IDUE is approximated to be independent of the unknown true frequencies, and the number of variables and privacy constraints are $2t$ and $t^2$ respectively ($t$ is the number of privacy levels). Note that the MSE of the naive mechanism without encoding (discussed in Sec. \ref{sec:challenges})  does not have closed-form expression and is dependent on the unknown true frequencies (thus the objective function cannot be directly evaluated), and the corresponding optimization problem  has $t^2$ variables and $t^3$  constraints. 

The proposed mechanism IDUE  works well for single-item data. However, when the input is an \emph{item-set}, i.e., any subset of the item domain, solving the optimization problem to determine the perturbation probabilities is not scalable due to an exponential blowup of the number of subsets. Thus, we combine our IDUE mechanism with Padding-and-Sampling protocol \cite{wang2018locally} to design a novel IDUE-PS mechanism for set-valued data. The privacy budget of a set is a function of the individual privacy budgets of items in the set. We will show that the perturbation probabilities of IDUE-PS (for item-set input with an exponential blowup) can be determined by IDUE (for single-item input) to satisfy MinID-LDP (in Theorem \ref{thm:MinID-LDP}) with a scalable optimization problem. Given the privacy level of each input, our proposed mechanisms satisfying MinID-LDP provide better privacy-utility tradeoff than $\epsilon$-LDP (where $\epsilon=\min_{x\in\mathcal{D}}\{\epsilon_x\}$). It is because our mechanisms achieve fine-grained privacy protection; whereas, the existing mechanisms satisfying LDP guarantee the highest privacy level.

Main contributions are summarized as follows:

(1) We introduce a new privacy notion called Input-Discriminative LDP (ID-LDP) with an instantiation termed MinID-LDP, which allows  finer-grained protection for different inputs than LDP.

(2) We design the Input-Discriminative Unary Encoding (IDUE) mechanism for single-item input that satisfies MinID-LDP and propose the frequency estimation protocol with an unbiased estimator. To minimize the Mean Squared Error (MSE) of IDUE, we formulate an optimization problem to solve the perturbation probabilities for the mechanism and derive three practical variants of the optimization model. 

(3) We extend IDUE with Padding-and-Sampling into IDUE-PS  for frequency estimation of item-set data, and show that it satisfies MinID-LDP with the same computation cost as IDUE that is designed for single-item input.

(4) We validate the correctness of the theoretical MSE analysis and effectiveness of our notion and mechanisms on synthetic and real-world datasets with both single-item and item-set types of input. We show that the proposed mechanisms outperform the existing ones for frequency estimation on categorical data. Also, the advantage of our mechanisms under the notion of MinID-LDP is enhanced when the distribution of privacy budgets of all inputs are more skewed.

\section{Related Work}

The notion of differential privacy (DP) in centralized setting was first introduced by Dwork in \cite{dwork2006differential}. It assumes a trusted server that possesses all genuine dataset. Then, a number of variants of differential privacy have been studied to provide different types of privacy guarantees such as $d$-privacy \cite{chatzikokolakis2013broadening}, Pufferfish privacy \cite{kifer2012rigorous}, 
Blowfish privacy \cite{he2014blowfish}, Concentrated DP \cite{bun2016concentrated}, and Personalized DP \cite{jorgensen2015conservative}. On the other hand, Duchi et al. \cite{duchi2013local} studied local differential privacy (LDP) without the assumption of a trusted server, and many mechanisms are proposed to applied to diverse data types/tasks, such as frequency estimation \cite{erlingsson2014rappor,wang2017locally}, set-valued data \cite{wang2018locally}, and key-value data \cite{ye2019privkv,gu2019pckv}. Several variants of LDP and the corresponding mechanisms have been studied, e.g., Personalized LDP \cite{nie2019utility,chen2016private}, Geo-indistinguishability \cite{andres2013geo,wang2017local,gu2019supporting},  Condensed LDP \cite{gursoy2019secure}  and  Utility-optimized LDP \cite{murakami2019utility}. We will compare these notions with the proposed one in Sec. \ref{sec:comparison_notions}. 

\section{Problem Statement and Preliminaries}
\label{sec:preliminaries}

\subsection{Problem Statement}
\label{sec:prob_state}
\textbf{System Model.} 
Our system model involves one data server and $n$ users $\mathcal{U}=\{u_1,u_2,\cdots,u_n\}$. Each user possesses one item or item-set in an item universe $\mathcal{I}=\{1,2,\cdots,m\}$  and perturbs it independently via a random perturbation mechanism before uploading it to the server. Then, the server collects users' data and computes the statistical information of users' data (we focus on frequency estimation in this paper). We consider two types of input data, one is the single-item input with domain $\mathcal{I}$, where each user only  possesses one item from $\mathcal{I}$; another is the item-set input with domain $\mathcal{P}(\mathcal{I})\triangleq\{x|x\subseteq\mathcal{I}\}$ (i.e., the power set of $\mathcal{I}$ with size $2^m$), where each user can possess any subset of $\mathcal{I}$. Assume there are $t$ privacy levels, and the $i$-th level only contains a subset of  $\mathcal{I}$, denoted by $\mathcal{I}_i$. Though the domain of the items can be large, the number of privacy levels determined by categories is usually small in practice, hence the usability and scalability of the system are guaranteed. For convenience, we denote the set of privacy budgets of all items in $\mathcal{I}$ as $\mathcal{E}=\{\epsilon_i\}_{i\in\mathcal{I}}$.

\textbf{Threat Model.}
We assume the server is untrusted, and each user only trusts herself, because data stored on the server can be revealed via either hacking activities or due to the server selling the data to a third party. 
Therefore, the adversary is assumed to possess the uploaded (perturbed) data of all users and it also knows the perturbation mechanism and the privacy budgets for all the inputs.

\textbf{Utility of Frequency Estimation.} 
The true frequency of an item $i\in\mathcal{I}$ is defined as the number of users who possess $i$
\begin{align}
\label{equ:c*}
c^{*}_i = \sum\nolimits_{u\in\mathcal{U}}\mathbbm{1}_{x_u}(i)
\quad(\forall i\in\mathcal{I})
\end{align}
where $x_u$	is the raw (input) data  of a user $u\in\mathcal{U}$ and can be a single-item or an item-set depending on the application scenario, and $\mathbbm{1}_{x_u}(i)$ is the indicator function, which is equal to 1 if $i\in x_u$ and equal to 0 otherwise. Note that $i$ only denotes one item from $\mathcal{I}$, while $x_u$ can be a subset of $\mathcal{I}$. After collecting the perturbed (output) data from all users, the server can estimate the frequency of an item $i\in\mathcal{I}$ via an estimator $\hat{c}_i$, which is a function of the perturbed data $\{y_u\}_{u\in\mathcal{U}}$ and mechanism parameters. The utility of frequency estimation is defined by the total Mean Squared Error (MSE) of estimators, i.e., $\text{MSE}=\sum\nolimits_{i=1}^{m}\text{MSE}_{\hat{c}_i}$, which will be minimized in the design of mechanism with privacy constraints.

\subsection{The Notion of LDP}
\label{sec:LDP_notion}


In the local setting, each user independently perturbs her input $x$ (raw data) using a mechanism $\mathcal{M}$ and uploads $\mathcal{M}(x)$ to the server for data analysis.
\begin{definition}[Local Differential Privacy (LDP) \cite{duchi2013local}]
	For a given $\epsilon\in\mathbb{R}^{+}$, a randomized mechanism $\mathcal{M}$ satisfies $\epsilon$-LDP if and only if for any pair of inputs $x,x^{\prime}$ and any output $y$
	\begin{align}
	\frac{\Pr(\mathcal{M}(x)=y)}{\Pr(\mathcal{M}(x^{\prime})=y)} \leqslant e^{\epsilon}
	\end{align}
\end{definition}

Intuitively, given an output $y$ of a mechanism $\mathcal{M}$, an adversary cannot infer with high confidence (controlled by $\epsilon$) whether the input is $x$ or $x^{\prime}$, which provides plausible deniability for individuals involved in the sensitive data. Here, $\epsilon$ is a parameter called \emph{privacy budget} that controls the strength of privacy protection. A smaller $\epsilon$ indicates stronger privacy protection because the adversary has lower confidence when trying to distinguish any pair of inputs $x,x^{\prime}$. LDP has the property of sequential composition, which guarantees the overall privacy for a sequence of mechanisms that satisfy LDP. 
\begin{theorem}[Sequential Composition of LDP \cite{mcsherry2009privacy}]
	If randomized mechanism $\mathcal{M}_i: \mathcal{D}\rightarrow\mathcal{R}_i$ satisfies $\epsilon_i$-LDP for $i=1,2,\cdots,k$, then their sequential combination $\mathcal{M}: \mathcal{D}\rightarrow\mathcal{R}_1\times\mathcal{R}_2\times\cdots\times\mathcal{R}_k$ defined by $\mathcal{M}=(\mathcal{M}_1,\mathcal{M}_2,\cdots,\mathcal{M}_k)$ satisfies $(\sum_{i=1}^k\epsilon_i)$-LDP.
\end{theorem}
According to sequential composition, a given privacy budget $\epsilon$ can be split into multiple portions, where each portion corresponds to the privacy budget of a randomized mechanism. 

\subsection{Mechanisms Satisfying LDP}
\label{sec:LDP_mechanisms}

\textbf{Randomized Response.} Randomized Response (RR) \cite{warner1965randomized} is a technique developed for the participants in a survey to return a randomized answer to a sensitive question to protect their privacy. Specifically, each participant gives a genuine answer with probability $p$ or gives the opposite answer with probability $1-p$, where $p=\frac{e^\epsilon}{e^\epsilon+1}$ to satisfy $\epsilon$-LDP. The standard RR only works for binary data (yes-or-no answers), but it can be extended to apply to $m$ categories by Generalized Randomized Response or  Unary Encoding. 

\textbf{Generalized Randomized Response.} 
The perturbation function in Generalized Randomized Response (GRR) \cite{wang2018locally} is 
\begin{align*}
\Pr(\mathcal{M}(x)=y)=
\begin{cases}
p, & \text{if } y=x\\
q, & \text{if } y\neq x
\end{cases},
\quad(\forall x,y=1,2,\cdots,m)
\end{align*}
To satisfy $\epsilon$-LDP, the probabilities are
$p=\frac{e^\epsilon}{e^\epsilon+m-1}$ and $q=\frac{1}{e^\epsilon+m-1}$, both of which would be small when the domain size $m$ is very large compared with $e^{\epsilon}$.

\textbf{Unary Encoding.} 
Unary Encoding (UE) \cite{wang2017locally} converts the input $x=i$ into a vector $\mathbf{x}=[0,\cdots,0,1,0,\cdots,0]$ with length $m$ where only the $i$-th bit is 1. Then each user perturbs each bit of  $\mathbf{x}$ independently with the following probabilities
\begin{align*}
\Pr(\mathbf{y}[k]=1)=
\begin{cases}
p, & \text{if } \mathbf{x}[k]=1\\
q, & \text{if } \mathbf{x}[k]=0
\end{cases}\quad
(\forall k=1,2,\cdots,m)
\end{align*}
This mechanism satisfies LDP with $\epsilon=\ln\frac{p(1-q)}{(1-p)q}$ \cite{wang2017locally}. The selection of $p$ and $q$ under a given privacy budget $\epsilon$ varies for different mechanisms. For example, the basic RAPPOR \cite{erlingsson2014rappor} assigns $p=\frac{e^{\epsilon/2}}{e^{\epsilon/2}+1}, q=1-p$, while the Optimized Unary Encoding (OUE) \cite{wang2017locally} assigns $p=0.5,
q=\frac{1}{e^{\epsilon}+1}$, which are obtained by optimizing the approximate variance.

\textbf{Frequency Estimation for GRR, RAPPOR and OUE.}
After receiving the perturbed data from all users, the server can implement the summation to get the total count of each bit, denoted by $c_i$ for the $i$-th bit. Then, the server calibrates the collected counts by an unbiased estimator $\hat{c}_i=\frac{c_i-nq}{p-q}$, whose Mean Squared Error (MSE) is equal to its variance \cite{wang2017locally}
\begin{align*}
\text{MSE}_{\hat{c}_i}=\text{Var}[\hat{c}_i]=\frac{nq(1-q)}{(p-q)^2}+\frac{c_i^{*}(1-p-q)}{p-q}
\end{align*}
where $c_i^{*}$ is the ground truth of the counting for item $i$. In summary, OUE can provide higher utility than RAPPOR for frequency estimation under the same $\epsilon$ due to the optimization, and the utility of GRR would be deteriorated much more than the other two mechanisms when the domain size $m$ is large.

\section{Input-Discriminative LDP}

In this section, a new privacy notion called ID-LDP is  introduced, which can provide input-discriminative protection with LDP. In ID-LDP, the indistinguishability level of a pair of possible inputs $x,x^{\prime}$ is determined by the corresponding privacy budgets $\epsilon_x,\epsilon_{x^{\prime}}$ of both inputs. Then, one instantiation of ID-LDP called MinID-LDP is formalized. It is proven to satisfy sequential composition theorem, which is an important property to guarantee the overall privacy for multiple query functions sequentially applied to the same database. Finally, our notion is compared with several existing privacy notions in the local setting and their relations are discussed.

\subsection{Definition}
\label{sec:definition}
LDP defines privacy as the maximum level of indistinguishability between any two possible inputs. In practical applications, the privacy levels of different inputs could be distinct. Thus, the requirement of indistinguishability between different pairs of inputs could be diverse. However, LDP cannot provide such fine-grained privacy protection because its definition is based on the worst-case scenario. Intuitively, discriminating inputs with different privacy levels and providing distinct protection to them can improve the utility of the query service due to the fine-grained protection for different inputs. We define the new notion ID-LDP as follows.

\begin{definition}[Input-Discriminative LDP (ID-LDP)]
	\label{def:ID-LDP}
	For a given privacy budget set $\mathcal{E}=\{\epsilon_x\}_{x\in\mathcal{D}}\in\mathbb{R}_{+}^{|\mathcal{D}|}$, where $|\mathcal{D}|$ is the size of the input domain $\mathcal{D}$, the randomized mechanism $\mathcal{M}$ satisfies $\mathcal{E}$-ID-LDP if and only if for any pair of inputs $x,x^{\prime}\in\mathcal{D}$, and any output $y\in Range(\mathcal{M})$
	\begin{align}
	\label{equ:ID-LDP}
	\frac{\Pr(\mathcal{M}(x)=y)}{\Pr(\mathcal{M}(x^{\prime})=y)} \leqslant e^{r(\epsilon_{x},\epsilon_{x^{\prime}})}
	\end{align}
	where $r(\cdot,\cdot)$ is a function of two privacy budgets.
\end{definition}

In Definition \ref{def:ID-LDP}, we assume inputs $x$ and $x^{\prime}$ belong to different privacy levels with privacy budgets $\epsilon_x$ and $\epsilon_{x^{\prime}}$ respectively and introduce a system-defined function $r(\epsilon_x,\epsilon_{x^{\prime}})$ to quantify the indistinguishability between $x$ and $x^{\prime}$. Note that the value of $\epsilon_x$ for each input $x$ is not sensitive information because  $\epsilon_x$ is independent of the users' raw data. In this paper, we assume  $\{\epsilon_x\}_{x\in\mathcal{D}}$ are universally set by the service provider. Note that, our notion can be easily combined with personalized LDP (PLDP) to reflect different privacy preferences of different users, in which case the privacy levels of all inputs can be set by users themselves. Theoretically, the notion of ID-LDP does not restrict the data type, which means it can be applied for categorical data, numerical data, or even the hybrid with multi-dimensions. In this paper, we mainly study the mechanism that satisfies ID-LDP for categorical data (single-item or item-set).

ID-LDP can provide input-discriminative protection with the function $r(\cdot,\cdot)$. In this paper, we mainly consider the minimum function between $\epsilon_{x}$ and $\epsilon_{x^{\prime}}$ as the privacy budget of a pair of inputs $x,x^{\prime}$, formulated by the following definition.

\begin{definition}[MinID-LDP]
	A randomized mechanism $\mathcal{M}$ satisfies $\mathcal{E}$-MinID-LDP if and only if it satisfies $\mathcal{E}$-ID-LDP with $r(\epsilon_{x},\epsilon_{x^{\prime}})=\min\{\epsilon_{x},\epsilon_{x^{\prime}}\}$.
\end{definition}

Intuitively, for any pair of inputs $x,x^{\prime}$, MinID-LDP guarantees that the adversary's capability of distinguishing $x$ and $x^{\prime}$ would not exceed the bound controlled by both $\epsilon_x$ and $\epsilon_{x^{\prime}}$, which achieves the worse-case privacy like LDP but \emph{only for the pair}. We use an example to show the benefit of our notion.  

\textbf{Example.}
Assume a health organization is taking a survey which asks $n$ participants to return a response perturbed from categories \{HIV, anemia, headache, stomachache, toothache\}, indexed by an integer $i$ from \{1,2,3,4,5\}. Since  HIV ($i=1$) is more sensitive than the others, the privacy budget that represents the privacy level should be different, such as $\epsilon_1=\ln4$ for HIV and $\epsilon_i=\ln6~(i\neq 1)$ for the others, where a smaller $\epsilon$ indicates a higher privacy level that needs stronger privacy protection. To satisfy LDP, all categories will be perturbed under the privacy budget $\epsilon_1=\ln 4$, even though some of them (such as anemia and headache) do not need such strong privacy protection. Under MinID-LDP, however, anemia and headache can be perturbed with less noise as long as the indistinguishability of any pair of inputs is bounded by both two budgets of them. We will compare the utility of mechanisms under the two notions in Sec. \ref{sec:comparison_mechanisms}.

As mentioned in Sec. \ref{sec:LDP_notion}, sequential composition is an important property to guarantee the overall privacy for a sequence of mechanisms.  The following theorem shows that MinID-LDP satisfies sequential composition as well.

\begin{theorem}[Sequential Composition of MinID-LDP]
	\label{thm:composition_min}
	If randomized mechanism $\mathcal{M}_i: \mathcal{D}\rightarrow\mathcal{R}_i$ satisfies $\mathcal{E}_i$-MinID-LDP for $i=1,2,\cdots,k$, where $\mathcal{E}_i=\{\epsilon_x^{(i)}\}_{x\in\mathcal{D}}\in\mathbb{R}^{|\mathcal{D}|}_{+}$, then their combination $\mathcal{M}: \mathcal{D}\rightarrow\mathcal{R}_1\times\mathcal{R}_2\times\cdots\times\mathcal{R}_k$ defined by $\mathcal{M}=(\mathcal{M}_1,\mathcal{M}_2,\cdots,\mathcal{M}_k)$ satisfies $(\sum_{i=1}^k\mathcal{E}_i)$-MinID-LDP, where $(\sum_{i=1}^k\mathcal{E}_i)\triangleq\{\sum_{i=1}^k\epsilon_x^{(i)}\}_{x\in\mathcal{D}}$.
\end{theorem}
\begin{IEEEproof}
	Let $x,x^{\prime}\in\mathcal{D}$ be any pair of inputs, for any output $y=(y_1,y_2,\cdots,y_k)\in\mathcal{R}_1\times\mathcal{R}_2\times\cdots\times\mathcal{R}_k$, we have
	\begin{align*}
	\frac{\Pr(\mathcal{M}(x)=y)}{\Pr(\mathcal{M}(x^{\prime})=y)}
	&= \prod_{i=1}^{k}\frac{\Pr(\mathcal{M}_i(x)=y_i)}{\Pr(\mathcal{M}_i(x^{\prime})=y_i)}
	\leqslant\prod_{i=1}^{k}e^{\min\{\epsilon_{x}^{(i)},\epsilon_{x^{\prime}}^{(i)}\}}\\
	&\leqslant\prod\nolimits_{i=1}^{k} e^{\epsilon_{x}^{(i)}}
	=e^{\sum_{i=1}^k\epsilon_{x}^{(i)}}
	\end{align*}
	Similarly, $\frac{\Pr(\mathcal{M}(x)=y)}{\Pr(\mathcal{M}(x^{\prime})=y)}\leqslant e^{\sum_{i=1}^k\epsilon_{x^{\prime}}^{(i)}}$. Finally, we have
	\begin{align*}
	\frac{\Pr(\mathcal{M}(x)=y)}{\Pr(\mathcal{M}(x^{\prime})=y)}
	\leqslant e^{\min\left\{\sum\nolimits_{i=1}^k\epsilon_{x}^{(i)},\sum\nolimits_{i=1}^k\epsilon_{x^{\prime}}^{(i)}\right\}}
	\end{align*}
	which indicates that $\mathcal{M}$ satisfies $(\sum_{i=1}^k\mathcal{E}_i)$-MinID-LDP.
\end{IEEEproof}

\subsection{Relationships and Comparison with Other Notions}
\label{sec:comparison_notions}

\textbf{Relationships with LDP.} 
If the privacy budgets for all inputs are the same, i.e., $\epsilon_{x}=\epsilon$ for all $x\in\mathcal{D}$, then $\mathcal{E}$-MinID-LDP  becomes $\epsilon$-LDP, which means MinID-LDP is a generalized version of LDP. In general, we have the following lemma to show their relationships.

\begin{lemma}
\label{lem:relationship}
If a mechanism satisfies $\epsilon$-LDP, then it also satisfies $\mathcal{E}$-MinID-LDP for all $\mathcal{E}$ with $\min\{\mathcal{E}\}=\epsilon$. On the other hand, if a mechanism satisfies $\mathcal{E}$-MinID-LDP, then it also satisfies $\epsilon$-LDP, where $\epsilon=\min\{\max\{\mathcal{E}\}, 2\min\{\mathcal{E}\}\}$.
\end{lemma}
\begin{IEEEproof}
    First, the following property can be directly derived from the definitions of LDP and MinID-LDP
    \begin{align*}
    \min\{\mathcal{E}\}\text{-LDP}\Rightarrow\mathcal{E}\text{-MinID-LDP}\Rightarrow\max\{\mathcal{E}\}\text{-LDP}
    \end{align*}
    Therefore, we only need to show that $\mathcal{E}$-MinID-LDP also implies $2\min\{\mathcal{E}\}$-LDP. Denote $x^{*}$ as the input that has the minimum budget, i.e., $\epsilon_{x^{*}}=\min\{\mathcal{E}\}$. Then, for all $x$, $x^{\prime}$ and $y$, the following inequality is satisfied under $\mathcal{E}$-MinID-LDP
    \begin{align*}
        \frac{\Pr(\mathcal{M}(x)=y)}{\Pr(\mathcal{M}(x^{\prime})=y)}
        &=\frac{\Pr(\mathcal{M}(x)=y)}{\Pr(\mathcal{M}(x^{*})=y)}\cdot\frac{\Pr(\mathcal{M}(x^{*})=y)}{\Pr(\mathcal{M}(x^{\prime})=y)}\\
        &\leqslant e^{\epsilon_{x^{*}}}\cdot e^{\epsilon_{x^{*}}}
        =e^{2\epsilon_{x^{*}}}=e^{2\min\{\mathcal{E}\}}
    \end{align*}
    which means $\mathcal{E}$-MinID-LDP implies $2\min\{\mathcal{E}\}$-LDP.
\end{IEEEproof}

From Lemma \ref{lem:relationship}, MinID-LDP relaxes LDP in at most twice of the privacy budget $\epsilon=\min\{\mathcal{E}\}$. It is due to the symmetric property of the  indistinguishability definition, so in a fully-connected policy graph,  if we require every pair of inputs $x,x^{\prime}$ to be indistinguishable with $\min\{\epsilon_x, \epsilon_{x^{\prime}}\}$, transitivity of indistinguishability yields $2\min\{\mathcal{E}\}$  between any pair of inputs. Note that the twice relaxation in privacy budget does not mean utility improvement is at most twice compared to LDP (depending on the query and data distribution). Although MinID-LDP can be regarded as a relaxation compared with LDP, in practice users' privacy expectation is naturally  different for different inputs, hence our notion captures user's fine-grained requirement, while LDP is too strong (i.e., provides overprotection) in this regard.

\begin{figure}[t!]
	\centering
	\includegraphics[width=3.4in]{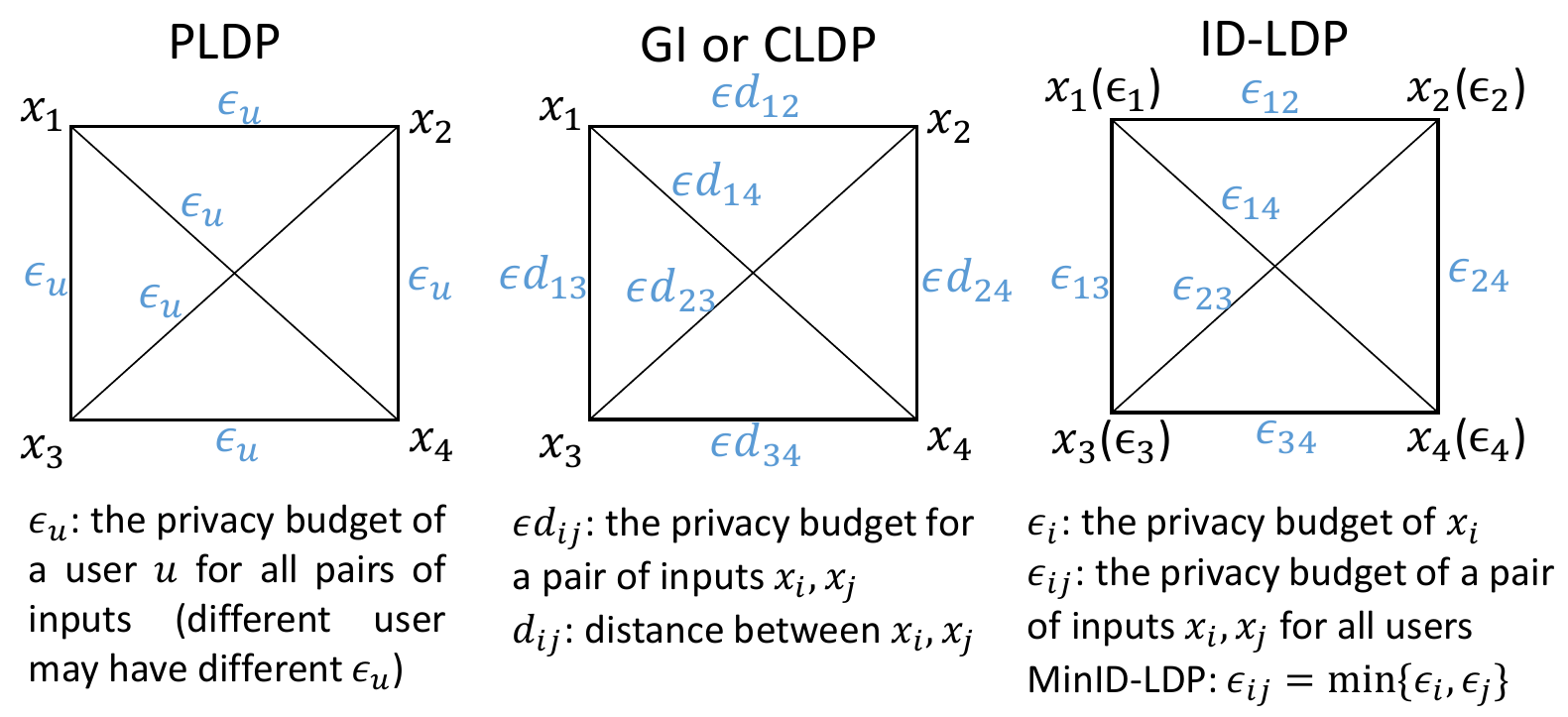}
	\vspace{-3mm}
	\caption{Privacy budget of a pair of inputs in several notions.}
	\label{fig:definitions}
	\vspace{-4mm}
\end{figure}

\textbf{Related Privacy Notions.}
LDP provides the worst-case privacy protection for all users and all inputs, where the global privacy budget is $\epsilon=\min_{x\in\mathcal{D}}\{\epsilon_x\}$. Several variants of LDP are related to our notion, but they have different ideas.  Fig. \ref{fig:definitions} shows the differences of personalized LDP (PLDP) \cite{nie2019utility},  geo-indistinguishability (GI) \cite{andres2013geo}, condensed LDP (CLDP) \cite{gursoy2019secure}, and our notion ID-LDP, in the form of their privacy policies where the vertices are inputs and edges are the distinguishability level (represented by privacy budget) of each pair of inputs. PLDP in \cite{nie2019utility,wang2015personalized} provides user-discriminative privacy requirements, i.e., each user can have personalized privacy budget which is often assumed to be unrelated to the raw data if it would be published. Another PLDP notion \cite{chen2016private} considers both safe region and privacy budget for each user in location-based systems. In summary, PLDP provides different protections for different users but does not differentiate different pairs of inputs. On the other hand, geo-indistinguishability  \cite{andres2013geo} in location-based systems and CLDP \cite{gursoy2019secure} in the data collection setting can provide distance-discriminative privacy, but they originate from an input pair-centric viewpoint and requires a distance metric for the inputs, where the distance metric (satisfying triangle inequality) may be hard to define for some data types such as categorical data. In contrast, our notion ID-LDP provides input-discriminative privacy requirements, where each input has a privacy budget (inputs with the same privacy level have the same budget), and the distinguishability of a pair of inputs can be determined by a function of the budgets of the two inputs to bound the distinguishability of this pair. Another notion that also considers distinct privacy levels is Utility-optimized LDP (ULDP) \cite{murakami2019utility}, which provides a privacy guarantee equivalent to LDP only for sensitive data to add less noise and improve utility. It can be regarded as a special case of the proposed MinID-LDP under two privacy levels of inputs (sensitive and non-sensitive) but with incomplete privacy policy graph, where sensitive and non-sensitive inputs can be fully distinguished when observing some outputs (termed invertible data) that reveals non-sensitive inputs, thus ULDP does not guarantee LDP. However, our MinID-LDP relaxes LDP by providing distinct bounds of privacy leakage for multiple (more than two) privacy levels  of inputs and also guarantees LDP with some privacy budgets (refer to Lemma \ref{lem:relationship}).

\setlength{\textfloatsep}{10pt}
\begin{table}[t!]
	\centering
	\small
	\caption{The bounds of prior-posterior $\frac{\Pr(x)}{\Pr(x|y)}~(\forall x,y)$.}
	\label{tab:leakage}
	\vspace{-2mm}
	\begin{tabular}{c|cc}
		\hline
		Privacy Notions & Lower Bound & Upper Bound\\
		\hline
		LDP & $e^{-\epsilon}$ & $e^{\epsilon}$\\
		PLDP & $e^{-\epsilon_u}$ & $e^{\epsilon_u}$\\
		GI or CLDP & $\sum_{x^{\prime}}\Pr(x^{\prime})e^{-\epsilon\cdot d(x,x^{\prime})}$ & $\sum_{x^{\prime}}\Pr(x^{\prime})e^{\epsilon\cdot d(x,x^{\prime})}$\\
		MinID-LDP & $e^{-\min\{\epsilon_x,2\min\{\mathcal{E}\}\}}$ & $e^{\min\{\epsilon_x,2\min\{\mathcal{E}\}\}}$\\
		\hline
	\end{tabular}
\end{table}

\textbf{Prior-Posterior Privacy Leakage Analysis.}
To understand our privacy notions in another perspective, we compare the prior-posterior privacy leakage (i.e., Local Information Privacy \cite{jiang2018context}) of the above notions.  Denote $\Pr(y|x)$ as the probability of outputting $y$ by given input $x$.  The ratio between the prior probability $\Pr(x)$ of an input $x$ and the posterior probability $\Pr(x|y)$ by observing the output $y$ can be computed as 
\begin{align}
\label{equ:leakage}
\frac{\Pr(x)}{\Pr(x|y)}=\frac{\Pr(y)}{\Pr(y|x)}
=\frac{\sum_{x^{\prime}\in\mathcal{D}}\Pr(x^{\prime})\Pr(y|x^{\prime})}{\Pr(y|x)}
\end{align}
which quantifies the privacy leakage  that the additional information the adversary can infer about an input $x$ by observing the output $y$. Note that (\ref{equ:leakage}) is different from Mutual Information (MI) \cite{cover2012elements} that quantifies the average leakage for all inputs and outputs. In our case, we only evaluate the bound of privacy leakage for a given input $x$ with an arbitrary output $y$. For different privacy notions, the lower bound and upper bound (independent of $y$) of prior-posterior privacy leakage defined by (\ref{equ:leakage}) are summarized in Table \ref{tab:leakage} (can be directly derived from definitions or Lemma \ref{lem:relationship}). The notion of LDP, PLDP, and MinID-LDP have the similar bound of privacy leakage for input $x$ with respect to the budget (though MinID-LDP has an additional bound with respect to $2\min\{\mathcal{E}\}$). However, LDP and PLDP do not differentiate the inputs, thus the budget would be assigned as the minimum value of all budgets in order to satisfy the privacy, but MinID-LDP can assign the required budget for different inputs, where the bound of leakage is also input-discriminative.

\section{Perturbation Mechanism and Frequency Estimation for Single-Item Input}
\label{sec:mechanism_single}
In this section, the considered input domain is $\mathcal{D}=\mathcal{I}$, i.e., single-item input. First, we formulate the optimization problem for designing a perturbation mechanism to optimize the utility of the frequency estimation of the outputs while satisfying MinID-LDP and the challenges to solve the problem. To address the challenges, we propose Input-Discriminative Unary Encoding (IDUE) mechanism and the corresponding unbiased frequency estimator.  Finally, we develop three practical variants of optimization model to obtain the optimal (or near-optimal) perturbation probabilities in IDUE.

\subsection{Objectives and Challenges}
\label{sec:challenges}
Our goal is to design a framework with perturbation mechanism and frequency estimation protocol that satisfies the proposed notion ID-LDP (MinID-LDP specifically) with the optimal Mean Squared Error (MSE) of frequency estimation. The general optimization problem can be modeled as 
\begin{align*}
\min~\text{MSE}, \quad s.t.\quad
\frac{\Pr(y|x)}{\Pr(y|x^{\prime})}
\leqslant e^{r(\epsilon_x,\epsilon_{x^{\prime}})}\quad(\forall x,x^{\prime}, y)
\end{align*}
However, solving this optimization problem has two challenges. First, the objective function cannot be directly evaluated in general because MSE is dependent on the unknown true frequencies. Second, the computation complexity is high since the numbers of variables (perturbation parameters/probabilities that determine the ratio $\frac{\Pr(y|x)}{\Pr(y|x^{\prime})}$ need to be solved) and privacy constraints (which should be satisfied for any inputs $x$, $x^{\prime}$ and output $y$) can be very large.

For example, a direct way to design such mechanism is to assign a perturbation/mapping matrix $\mathbf{P}\in\mathbb{R}^{|\mathcal{D}|\times|\mathcal{D}|}$ (which can be regarded as a variant of GRR discussed in Sec. \ref{sec:LDP_mechanisms}), where each element represents the perturbation probability $\Pr(y|x)$ for $x,y\in\mathcal{D}$ (the output domain  $\mathcal{R}=\mathcal{D}$ in this case). However, solving the elements in matrix $\mathbf{P}$ by minimizing MSE under the privacy constraints has several issues in practice. First, $\hat{\mathbf{c}}=(\mathbf{P}^{\text{T}})^{-1}\mathbf{c}$ was shown to be the unbiased estimator of the true frequency vector $\mathbf{c}^{*}$ \cite{huang2008optrr}, where $\mathbf{c}$ is the calculated frequency vector of outputs. But the elements in  the inversion matrix $(\mathbf{P}^{\text{T}})^{-1}$ do not have closed-form expression in general and the MSE of this estimation is dependent on the unknown true frequencies, thus the objective function of minimizing MSE cannot be directly evaluated. Note that the frequency estimator of the original GRR discussed in Sec. \ref{sec:LDP_mechanisms} can be regarded as a special case of the above mechanism, where the inversion matrix can be explicitly calculated (because there are only two different perturbation probabilities) and the term that is related to the true frequencies only takes a small portion of MSE, thus the \emph{approximate}  MSE can be independent of the unknown true frequencies. But it cannot be applied here because the perturbation probabilities for different inputs are designed to be different in our setting. Second, since the numbers of variables and constraints are $|\mathcal{D}|^2$ (all elements in $\mathbf{P}$) and $|\mathcal{D}|^3$ (for all $x,x^{\prime}$ and $y$) respectively, the computation cost would be very high especially for item-set input $\mathcal{D}=\mathcal{P}(\mathcal{I})$ with $|\mathcal{D}|=2^m$. Third, the domain size $|\mathcal{D}|$ is usually very large in practice, then the perturbation probabilities will become very small because of $\sum_{y\in\mathcal{D}}\Pr(y|x)=1$ for all $x\in \mathcal{D}$, which means the probability of reporting the true value is low, then the utility would greatly deteriorate. 

In the following, we propose the Unary Encoding (refer to Sec. \ref{sec:LDP_mechanisms}) based perturbation mechanism and frequency estimation protocol for single-item input with $\mathcal{D}=\mathcal{I}$ under privacy notion ID-LDP. Due to the nature of Unary Encoding, there are less perturbation parameters (shown in Sec.  \ref{sec:IDUE}), and the upper bound for all $y$ of ratio $\frac{\Pr(y|x)}{\Pr(y|x^{\prime})}$  when fixing $x$ and $x^{\prime}$ can be explicitly calculated. Then, the equivalent constraint in \eqref{equ:constraint} needs to be satisfied only for all inputs $x$ and $x^{\prime}$ (hence the number of constraints is reduced). On the other hand, the unbiased estimator $\hat{c}_i$ in \eqref{equ:c_hat} can be explicitly expressed by the perturbation probabilities, and its MSE  in \eqref{equ:MSE_c} can be composed by two terms, where only the second term is dependent on the true frequency $c_i^{*}$. Finally, we address the challenge due to the lack of true frequencies by developing three variants of optimization models in Sec. \ref{sec:optimization} to obtain the approximate total MSE which is independent of true frequencies.


\subsection{Mechanism Design}
\label{sec:IDUE}


\textbf{Input-Discriminative Unary Encoding (IDUE).}
We first encode the single-item input $x=i$ into a $m$-length vector 
\begin{align}
\label{equ:v_i}
\mathbf{x}=\mathbf{v}_i=[0,\cdots,0,1,0,\cdots,0]
\end{align}
where vector $\mathbf{x}$ denotes the encoded input, $\mathbf{v}_i$ denotes the vector whose $i$-th position is 1 and other positions are 0s. Then, each bit of the input vector $\mathbf{x}$ is perturbed into 0 or 1 independently to get the output vector $\mathbf{y}$ with probabilities
{
\begin{align*}
\Pr(\mathbf{y}[k]=1|\mathbf{x}[k]=1)=a_k,~
\Pr(\mathbf{y}[k]=0|\mathbf{x}[k]=1)=1-a_k\\
\Pr(\mathbf{y}[k]=1|\mathbf{x}[k]=0)=b_k,~
\Pr(\mathbf{y}[k]=0|\mathbf{x}[k]=0)=1-b_k
\end{align*} 
}where we assume $a_k> b_k~(\forall k\in\mathcal{I})$ in order to obtain a good utility. 
Compared with the original Unary Encoding protocol \cite{wang2017locally} discussed in Sec. \ref{sec:LDP_mechanisms}, where $p$ and $q$ correspond to $a_i$ and $b_i$ in our notation, IDUE assigns different perturbation probabilities for different bits, which is the key point to achieve input-discriminative protection.

For two different input vectors $\mathbf{v}_i$ (only the $i$-th bit is 1) and $\mathbf{v}_j$, the probability ratio of distinguishing the pair of $\mathbf{v}_i$ and $\mathbf{v}_j$ by observing the output vector $\mathbf{y}$ is
\begin{align*}
\frac{\Pr(\mathbf{y}|\mathbf{v}_i)}{\Pr(\mathbf{y}|\mathbf{v}_j)}
=\frac{\prod_{k=1}^{m}\Pr(\mathbf{y}[k]|\mathbf{v}_i)}{\prod_{k=1}^{m}\Pr(\mathbf{y}[k]|\mathbf{v}_j)}
=\frac{\Pr(\mathbf{y}[i]|\mathbf{v}_i)\Pr(\mathbf{y}[j]|\mathbf{v}_i)}{\Pr(\mathbf{y}[i]|\mathbf{v}_j)\Pr(\mathbf{y}[j]|\mathbf{v}_j)}
\end{align*}
Since $a_k>b_k~(\forall k\in\mathcal{I})$, we have
\begin{align*}
\frac{\Pr(\mathbf{y}[i]|\mathbf{v}_i)\Pr(\mathbf{y}[j]|\mathbf{v}_i)}{\Pr(\mathbf{y}[i]|\mathbf{v}_j)\Pr(\mathbf{y}[j]|\mathbf{v}_j)}
=\frac{(\frac{a_i}{b_i})^{\mathbf{y}[i]}(\frac{1-a_i}{1-b_i})^{1-\mathbf{y}[i]}}{(\frac{a_j}{b_j})^{\mathbf{y}[j]}(\frac{1-a_j}{1-b_j})^{1-\mathbf{y}[j]}}
\leqslant\frac{a_i(1-b_j)}{b_i(1-a_j)}
\end{align*}
where the  left side equals the right side   if and only if $\mathbf{y}[i]=1$ and $\mathbf{y}[j]=0$. Then, the privacy constraint in (\ref{equ:ID-LDP}) is
\begin{align}
\label{equ:constraint}
\frac{a_i(1-b_j)}{b_i(1-a_j)}
\leqslant e^{r(\epsilon_i,\epsilon_j)}\quad(\forall i,j\in\mathcal{I})
\end{align}
By converting the original privacy constraint  into  (\ref{equ:constraint}),  which is independent of $y$ thus has less number of constraints, we can reduce the computational complexity compared with the direct formulation described in Sec. \ref{sec:challenges}. 

To obtain the optimal perturbation probabilities for our IDUE mechanism,  we first develop the frequency estimator for IDUE, and evaluate the theoretical MSE of the estimator as a function of perturbation probabilities. Then we formalize the optimization problem with three variants by minimizing the MSE with the privacy constraints in (\ref{equ:constraint}).

\subsection{Frequency Estimation}
\label{sec:freq_est}

Denote the collected frequency of the $i$-th bit as $c_i= \sum_{u\in\mathcal{U}}\mathbf{y}_u[i]$, where $\mathbf{y}_u$ is the output vector of a user $u\in\mathcal{U}$. For frequency estimation, we utilize the following estimator 
\begin{align}
\label{equ:c_hat}
\hat{c}_i=\frac{c_i-nb_i}{a_i-b_i}
\end{align}
which can be shown as the  unbiased estimator of the true frequency $c_i^{*}$ defined in (\ref{equ:c*}).

\begin{theorem}[Unbiasedness Property]
	If $a_i\neq b_i~(\forall i\in\mathcal{I})$, then $\mathbb{E}[\hat{c}_i]=c_i^{*}$, where estimator $\hat{c}_i$ is defined in (\ref{equ:c_hat}).
\end{theorem}
\begin{IEEEproof}
	Since $\mathbb{E}[c_i]
	=c^{*}_ia_i + \sum_{k\neq i}c^{*}_kb_i
	=c^{*}_ia_i + (n-c^{*}_i)b_i$, then we have $\mathbb{E}[\hat{c}_i]
	=\frac{\mathbb{E}[c_i]-nb_i}{a_i-b_i}=c^{*}_i$, which means $\hat{c}_i$ is an unbiased estimator of $c^{*}_i$. 
\end{IEEEproof}

The frequency estimator in (\ref{equ:c_hat}) can be regarded as the generalized version of the estimator that is used for the original Unary Encoding (refer to Sec. \ref{sec:LDP_mechanisms}). Due to the unbiasedness of estimator $\hat{c}_i$, the MSE of $\hat{c}_i$ is equal to its variance
\begin{align}
\label{equ:MSE_c}
\text{MSE}_{\hat{c}_i}
&=\text{Var}[\hat{c}_i]
=\frac{c^{*}_i a_i(1-a_i) + (n-c^{*}_i)b_i(1-b_i)}{(a_i-b_i)^2}\notag\\
&=\frac{nb_i(1-b_i)}{(a_i-b_i)^2}+
\frac{c^{*}_i(1-a_i-b_i)}{a_i-b_i}
\end{align}
In Sec. \ref{sec:optimization}, the summation of $\text{MSE}_{\hat{c}_i}$ will be minimized with the privacy constraints of ID-LDP.

\begin{table*}[t!]
	\centering
	\small
	\caption{Utility comparison in the toy example, where $\epsilon_1=\ln4$ and $ \epsilon_i=\ln6~(i\neq 1)$.}
	\label{tab:var}
	\renewcommand\arraystretch{1.2}
	\begin{tabular}{|c|c|c|c|c|c|c|c|c|}
		\hline
		 \multirow{3}{*}{Mechanisms} & \multirow{3}{*}{Privacy Notions} & \multicolumn{4}{c|}{Probability of flipping the $i$-th bit} & \multicolumn{2}{c|}{Variance of frequency estimation} & \textbf{Total variance}\\
		\cline{3-9}
		 &&\multicolumn{2}{c|}{$1-a_i$ (if $\mathbf{x}[i]=1$)} & \multicolumn{2}{c|}{$b_i$ (if $\mathbf{x}[i]=0$)} & \multicolumn{2}{c|}{$\text{Var}[\hat{c}_i]$} & \multirow{2}{*}{$\sum_i\text{Var}[\hat{c}_i]$} \\ \cline{3-8} 
		& & $i=1$ & $i=2\sim5$ & $i=1$ & $i=2\sim5$ & $i=1$ & $i=2\sim5$& \\
		\hline
		RAPPOR \cite{erlingsson2014rappor} & LDP &  $0.33$ &  $0.33$ & $0.33$  & $0.33$ &  $2n$ &  $2n$ & $10n$  \\
		OUE \cite{wang2017locally} & LDP & $0.5$ & $0.5$ &  $0.2$ &  $0.2$ & $1.78n+c_i$ & $1.78n+c_i$ &$9.9n$\\
		IDUE & MinID-LDP& $0.41$ & $0.33$ &  $0.33$ & $0.28$ &  $3.27n+0.31c_i$ & $1.32n+0.13c_i$&$8.68n\sim 8.86n$\\
		\hline
	\end{tabular}	
    \vspace{-4mm}
\end{table*}

\subsection{Finding Optimal Perturbation  Probabilities}
\label{sec:optimization}

As described in Sec. \ref{sec:prob_state}, the input domain is divided into $t$ subsets $\mathcal{I}_1,\mathcal{I}_2,\cdots,\mathcal{I}_t$ with different privacy levels. Denote the number of items in subset $\mathcal{I}_i$ as $|\mathcal{I}_i|=m_i$ and the privacy budget is $\epsilon_i~(i=1,2,\cdots,t)$. We can assign the same parameters $a_i$ and $b_i$ for all items in $\mathcal{I}_i$. If $t=1$, i.e., all items in $\mathcal{I}$ have the same $\epsilon$, then this case reduces to the LDP setting. The MSE of subset $\mathcal{I}_i$ is calculated by
{\small
\begin{align*}
\text{MSE}_{\mathcal{I}_i}=\sum_{k\in\mathcal{I}_i}\text{MSE}_{\hat{c}_k}
=\frac{nm_ib_i(1-b_i)}{(a_i-b_i)^2}+
\frac{(1-a_i-b_i)}{a_i-b_i}\sum_{k\in\mathcal{I}_i}c^{*}_k
\end{align*}
}The expression of $\text{MSE}_{\mathcal{I}_i}$ is dependent on the true frequency $\sum_{k\in\mathcal{I}_i}c^{*}_k$, which is unknown in practice, thus cannot be established as the objective function for the optimization problem. Therefore, we propose three variants of the optimization model, named  \verb|opt0|, \verb|opt1|, and \verb|opt2|, to make the  objective function independent of the true frequencies.

\textbf{\texttt{opt0}: Optimization Model in the Worst-Case.}
Though $\text{MSE}_{\mathcal{I}_i}$ is dependent on the true frequencies, we have the following upper bound of the total MSE since $\sum_{k\in \mathcal{I}_i}c_k^{*}\leqslant n$ to get rid of the unknown true frequency $c_k^{*}$
\begin{align*}
\sum_{i=1}^{t}\text{MSE}_{\mathcal{I}_i}
\leqslant\sum_{i=1}^{t}\frac{nm_ib_i(1-b_i)}{(a_i-b_i)^2}+\max\left\{\frac{1-a_i-b_i}{a_i-b_i}\right\}\cdot n
\end{align*}
which can be regarded as the total MSE in the worst-case. Then, determining parameters $\{a_i,b_i\}_{i=1}^t$ is converted to minimizing the worst-case MSE
\begin{align}
\label{equ:opt_ab}
\min_{a,b}\quad & f\triangleq\sum_{i=1}^{t}\frac{m_ib_i(1-b_i)}{(a_i-b_i)^2}+\max\left\{\frac{1-a_i-b_i}{a_i-b_i}\right\} \\
s.t.\quad & \frac{a_i(1-b_j)}{b_i(1-a_j)}
\leqslant e^{r(\epsilon_i,\epsilon_j)}\quad(\forall i,j=1,2,\cdots,t)\notag\\
& 0<b_i<a_i<1\quad(\forall i=1,2,\cdots,t) \notag
\end{align}
where the scaling constant $n$ in the objective function is omitted since it does not change the result. Since the feasible region of optimization problem (\ref{equ:opt_ab})  contains the perturbation probabilities of RAPPOR and OUE, the solution solved by (\ref{equ:opt_ab}) will have less worst-case MSE than both RAPPOR and OUE.

It can be shown that the objective function in (\ref{equ:opt_ab}) is not convex in the feasible region. To address this,  in the following we consider two types of space reducing strategies, which are related to RAPPOR  and OUE  respectively.  They can be used to find near-optimal solutions with convex property and reduced complexity compared with the formulation in \eqref{equ:opt_ab}. Our idea is to further constrain the variables (which shrinks the feasible region), so that the optimization problem can be convex.


\textbf{\texttt{opt1}: Optimization Model Constrained with RAPPOR Structure.}  RAPPOR regards bit-0 and bit-1 equally thus $p+q=1$. We 
add the corresponding constraint  $a_i+b_i=1~(\forall i)$ in our optimization problem and represent $a_i,b_i$ as
\begin{align}
\label{equ:ab_opt1}
a_i=\frac{e^{\tau_i}}{e^{\tau_i}+1},\quad
b_i=\frac{1}{e^{\tau_i}+1}
\quad(i=1,2,\cdots,t)
\end{align}  
where $\tau_i>0~(\forall i)$. Then $\frac{1-a_i-b_i}{a_i-b_i}=0$ and  the total MSE is
\begin{align*}
\sum_{i=1}^{t}\text{MSE}_{\mathcal{I}_i}
=\sum_{i=1}^{t}\frac{nm_ib_i(1-b_i)}{(a_i-b_i)^2}
=n\sum_{i=1}^{t}\frac{m_ie^{\tau_i}}{(e^{\tau_i}-1)^2}
\end{align*}
with privacy constraints
\begin{align*}
\frac{a_i(1-b_j)}{b_i(1-a_j)}=e^{\tau_i+\tau_j}\leqslant e^{r(\epsilon_i,\epsilon_j)}
~~\Leftrightarrow~~ 
\tau_i+\tau_j\leqslant r(\epsilon_i,\epsilon_j)
\end{align*}
Therefore, we can get the following optimization problem
\begin{align}
\label{equ:opt_tau}
\min_{\tau_1,\cdots,\tau_t>0} \quad& f(\tau)\triangleq\sum_{i=1}^{t}\frac{m_ie^{\tau_i}}{(e^{\tau_i}-1)^2} \\
s.t. \quad& \tau_i+\tau_j\leqslant r(\epsilon_i,\epsilon_j)\quad(\forall i, j) \notag
\end{align}

\textbf{\texttt{opt2}: Optimization Model Constrained with OUE Structure.} OUE focuses on less noise of  bit-0 thus $p=0.5$. We add the additional constraints $a_i=0.5~(\forall i)$ and rewrite the privacy constraints in (\ref{equ:constraint}) as
\begin{align*}
\frac{a_i(1-b_j)}{b_i(1-a_j)}
=\frac{1-b_j}{b_i}\leqslant e^{r(\epsilon_i,\epsilon_j)}
~~\Leftrightarrow~~ 
e^{r(\epsilon_i,\epsilon_j)}\cdot b_i + b_j \geqslant 1
\end{align*}
Since $a_i=0.5$, we have $\frac{1-a_i-b_i}{a_i-b_i}=1~(\forall i)$, then the total MSE can be represented by
{\small
\begin{align*}
\sum_{i=1}^{t}\frac{nm_ib_i(1-b_i)}{(a_i-b_i)^2} + \sum_{i=1}^{t}\sum_{k\in\mathcal{I}_i}c^{*}_k
=\sum_{i=1}^{t}\frac{nm_ib_i(1-b_i)}{(0.5-b_i)^2} +\sum_{k\in\mathcal{I}}c^{*}_k
\end{align*}
}Therefore, we can obtain the following optimization problem (omit the scaling constant $n$ and the additive constant $\sum_{k}c^{*}_k$)
\begin{align}
\label{equ:opt_b}
\min_{0<b_i<0.5} \quad& f(b)\triangleq\sum_{i=1}^{t}\frac{m_ib_i(1-b_i)}{(0.5-b_i)^2} \\
s.t. \quad& e^{r(\epsilon_i,\epsilon_j)}\cdot b_i + b_j \geqslant 1\quad(\forall i, j) \notag
\end{align}

\textbf{Summary of Three Models.} \verb|opt0| with \emph{non-convex} objective function has $2t$ variables and $t^2$ \emph{non-linear} privacy constraints. Both \verb|opt1| and \verb|opt2| have $t$ variables and  $t^2$ \emph{linear} privacy constraints, and the Hessian matrices of their objective functions are positive-definite in the feasible region, thus they are \emph{convex} problems with lower computation complexity. In common cases that only need a small number of privacy levels (i.e.,  a smaller $t$), we can use \verb|opt0| to obtain the theoretically optimal solution with acceptable computation overhead. But if $t$ is very large, it would be better to use \verb|opt1| or \verb|opt2| to obtain the near-optimal solution in the shrunk feasible region.

\subsection{Comparison with LDP Mechanisms}
\label{sec:comparison_mechanisms}

In the example discussed in Sec. \ref{sec:definition}, all participants randomly perturb their true answers with a certain probability  to protect privacy. Specifically, each participant first generates a vector $\mathbf{x}$ with five bits, where only the position of the truth is 1 and other positions are 0s, then flips each bit with assigned probabilities (depending on the mechanisms) to generate the perturbed vector $\mathbf{y}$. Finally, the organization aggregates all perturbed vectors from $n$ participants and estimate the counts of these categories by the estimator $\hat{c}_i$. In Table  \ref{tab:var}, we show that our proposed mechanism IDUE (solved by \verb|opt0|) outperforms the state-of-the-art mechanisms (RAPPOR \cite{erlingsson2014rappor} and OUE \cite{wang2018locally}) under the given privacy levels of inputs, where a smaller total variance $\sum_i\text{Var}[\hat{c}_i]$ indicates a better utility (MSE is equal to the variance for an unbiased estimator). In IDUE, the flipping probabilities for $i=1$ and $i\neq 1$ are different due to the different privacy levels, while mechanisms satisfying LDP (e.g., RAPPOR and OUE) do not differentiate them.  By adjusting the flipping probabilities for different bits, IDUE can achieve the optimal utility with the required protection. The total variance $\sum_i\text{Var}[\hat{c}_i]$ of our mechanism IDUE is in a range because it depends on the distribution of true input data. We can see that the upper bound is still less than that of the existing mechanisms, indicating that our mechanism outperforms others even in the worst-case. For IDUE, the probability of flipping the bit for $i=1$ may be larger than that in other mechanisms because  $\frac{a_1(1-b_j)}{b_1(1-a_j)}=4=e^{\epsilon_1}~(\forall j)$ in RAPPOR and OUE, thus to allow smaller flipping probabilities (i.e., larger $\frac{1-b_j}{1-a_j}$) for $j\neq1$ under the privacy constraint $\frac{a_1(1-b_j)}{b_1(1-a_j)}\leqslant e^{\epsilon_1}$ in \eqref{equ:constraint}, IDUE needs to increase the flipping probability (hence a larger variance) for $i=1$ to decrease $\frac{a_1}{b_1}$. This property of IDUE leads to a larger variance for $i=1$, but smaller flipping probabilities and variance for $i\neq1$, then the overall utility is improved.

\section{Mechanism for Item-Set Input}
\label{sec:mechanism_set}

In this section, we consider the item-set input, where the input domain is $\mathcal{D}=\mathcal{P}(\mathcal{I})$, i.e., the power set of $\mathcal{I}$. If we directly apply the IDUE mechanism developed in Sec. \ref{sec:mechanism_single} for this case, each possible set will need to be assigned two perturbation probabilities (for bit-0 and bit-1), therefore  the computational cost of solving the optimization problem would be very high because the size of the input domain is $2^m$. In this section, we solve the scalability issue by extending the IDUE mechanism  with Padding-and-Sampling (PS) protocol to adapt to item-set input. The privacy analysis shows that if mechanism IDUE satisfies MinID-LDP, then the extended one IDUE-PS satisfies MinID-LDP as well. Thus, IDUE-PS has the same computational complexity as IDUE.


\subsection{The Padding-and-Sampling Protocol}
\label{sec:padding-and-sampling}

Assume the raw data of each user is a set of items, where the number of items in each set can be different. This problem is more challenging than the single-item input even under LDP notion because the user has more than one item, where each item would split privacy budget (reporting all items will lead to large noise in each item and thus bad utility of query). However, if adopting sampling technique to avoid budget splitting, the different number of items in each user makes the frequency estimation much harder because the sampling probability depends on the number of items of the user which should be kept private. A good solution to address the item-set type of input is the Padding-and-Sampling protocol \cite{wang2018locally}.


Algorithm \ref{alg:PS} shows the steps of Padding-and-Sampling protocol, where the item-set $x\in\mathcal{D}$ is padded by a dummy set $\mathcal{S}$ (or truncated) into a new set $x_p$ with a fixed length $\ell$ and only one item $x_s$ is randomly sampled from the padded set $x_p$. The fixed length $\ell$ is a system parameter which will affect the utility in some way (depending on the data distribution).  More details of selecting a good $\ell$ is discussed in \cite{wang2018locally}. We will discuss how to select $\ell$ empirically in Sec. \ref{sec:itemset_data}.

\subsection{Mechanism Design and Privacy Analysis}

\begin{algorithm}[!t] 
    \small
	\caption{Padding-and-Sampling (PS) \cite{wang2018locally}} 
	\setstretch{1.0}
	\label{alg:PS} 
	\begin{algorithmic}[1] 		
		\REQUIRE Item-set $x\in\mathcal{D}$  and  dummy set $\mathcal{S}=\{m+1,\cdots,m+\ell\}$.
		\ENSURE One item $x_{s}\in x\cup\mathcal{S}$ 
		\STATE Set the padded input $x_{p}\leftarrow x$
		\IF{$|x|<\ell$}
		\STATE Select $(\ell-|x|)$ dummy items with uniform random from $\mathcal{S}$ and add them into $x_p$
		\ELSIF{$|x|>\ell$} 
		\STATE Drop out $(|x|-\ell)$ items with uniform random from $x_p$
		\ENDIF	
		\STATE Sample one item $x_s$ with uniform random from $x_p$
	\end{algorithmic}
\end{algorithm}

\begin{figure}[!t]
	\vspace{-2mm}
	\centering
	\includegraphics[width=3.4in]{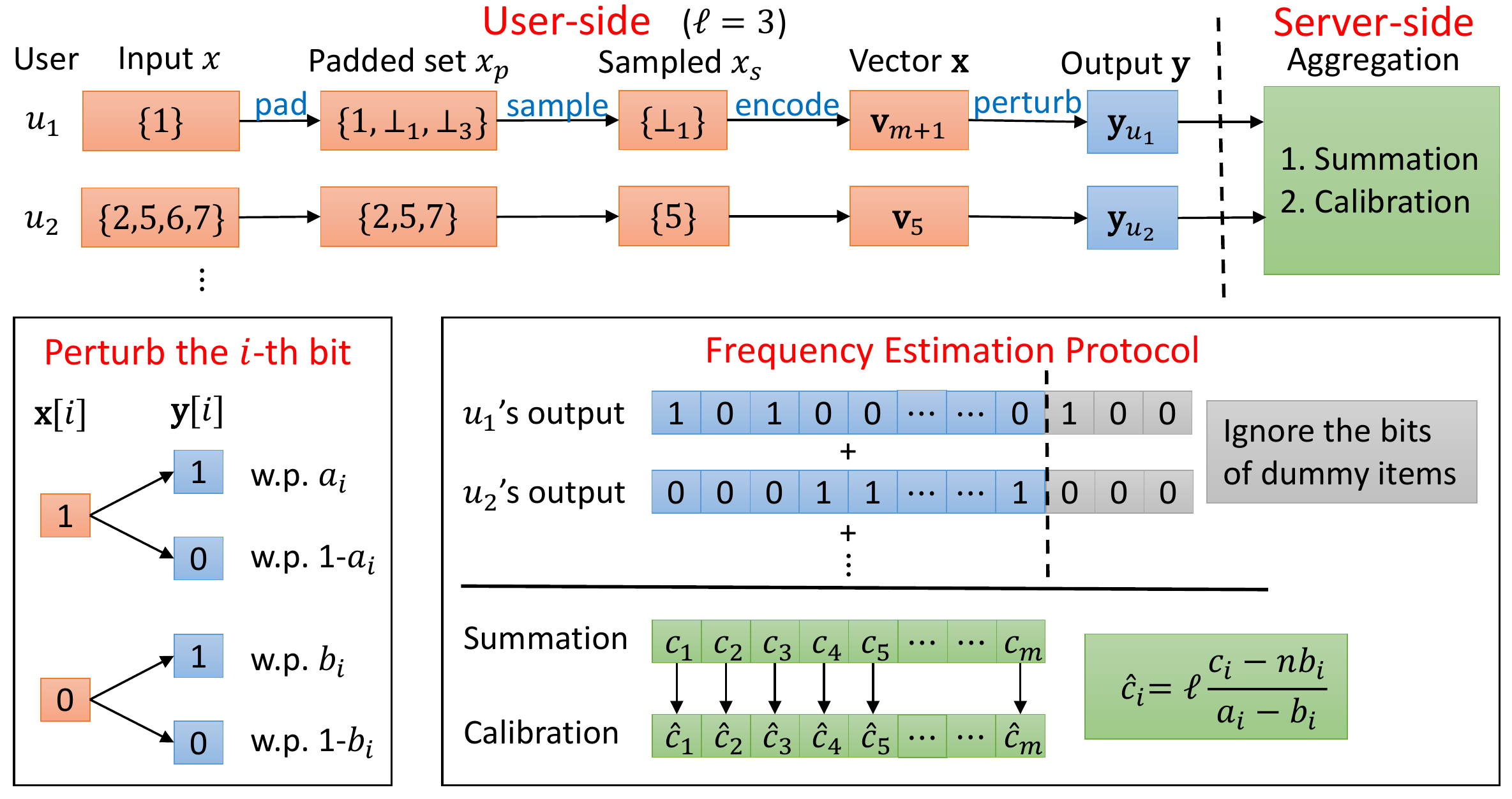}
	\vspace{-2mm}
	\caption{The IDUE-PS mechanism for item-set input.}
	\label{fig:protocol}
	\vspace{2mm}
\end{figure}

\textbf{IDUE with Padding-and-Sampling for Item-set Input.} 
By adopting the Padding-and-Sampling (PS) protocol, our previous mechanism IDUE (in Sec. \ref{sec:IDUE}) can be extended for set-valued input. Algorithm \ref{alg:IDUE-PS} shows the steps (sampling, encoding, and perturbing) of our extended mechanism named IDUE-PS, where the data is perturbed according to the sampled item's parameters under the single item case. Fig. \ref{fig:protocol} shows the  diagram of  perturbation steps in the user-side and aggregation (on frequency estimation) in the server-side. Since the original itemset input $x$ is padded with some dummy items from a domain $\mathcal{S}$ that is disjoint from the original item domain $\mathcal{I}$, the item domain is extended to be $\mathcal{I}\cup\mathcal{S}$. We denote the new item domain
$\mathcal{I}^{\prime}=\{1,2,\cdots,m+\ell\}$, where the last $\ell$ items are dummy items, and the encoded vector $\mathbf{x}$ has $(m+\ell)$ bits. Since each item will be sampled with probability $1/\ell$ from the padded set $x_p$, the result of frequency estimation  needs to be multiplied by the factor $\ell$, i.e., $\hat{c}_i=\ell\cdot\frac{c_i-nb_i}{a_i-b_i}$ for $i\in\mathcal{I}$ (we do not need to estimate frequencies of dummy items). 

Assume the perturbation probabilities of $i$-th bit are $a_i,b_i$, and denote two paramters
\begin{align}
\label{equ:alpha_beta}
\alpha_i=\frac{a_i}{b_i},\quad
\beta_i=\frac{1-a_i}{1-b_i}\quad(\forall i\in\mathcal{I}^{\prime})
\end{align}
Since $\alpha_i-\beta_i=\frac{a_i-b_i}{b_i(1-b_i)}$ and $0<b_i\leqslant a_i<1$, we have $1\leqslant\beta_i\leqslant\alpha_i$ ($\alpha_i=\beta_i$ only when $a_i=b_i$). 
Before proving the privacy guarantee of IDUE-PS, we show the following useful lemma first.

\begin{algorithm}[!t] 
	\small
	\caption{IDUE-PS for Item-Set Input} 
	\setstretch{1.0}
	\label{alg:IDUE-PS} 
	\begin{algorithmic}[1] 			
		\REQUIRE Item-set $x\in\mathcal{D}$ and dummy set $\mathcal{S}=\{m+1,\cdots,m+\ell\}$. Perturbation probabilities $(a_i,b_i)$ for $i\in\mathcal{I}^{\prime}=\{1,2,\cdots,m+\ell\}$.
		\ENSURE Vector $\mathbf{y}\in\{0,1\}^{m+\ell}$ 
		\STATE Let $\mathbf{x}
		=[0,\cdots,0]$ with length $(m+\ell)$
		\STATE Sample one item $x_{s}\in \mathcal{I}^{\prime}$ by Algorithm \ref{alg:PS} and let $\mathbf{x}[x_{s}]= 1$.
		\FOR{$k=1$ to $(m+\ell)$} 
		\IF{$\mathbf{x}[k]= 1$} 
		\STATE Randomly draw $\mathbf{y}[k]\sim\text{Bernoulli}(a_k)$
		\ELSE
		\STATE Randomly draw $\mathbf{y}[k]\sim\text{Bernoulli}(b_k)$
		\ENDIF
		\ENDFOR 
	\end{algorithmic}
\end{algorithm}

\begin{lemma}
	\label{lem:ratio}
	For any item-set inputs $x,x^{\prime}\in\mathcal{D}$, and any output $y$ of IDUE-PS (Algorithm \ref{alg:IDUE-PS}), the following probability ratio is bounded by
	\begin{align}
	\label{equ:ratio}
	\frac{\Pr(y|x)}{\Pr(y|x^{\prime})}
	\leqslant\frac{\eta_x\sum_{i\in x}\frac{\alpha_i}{|x|}+(1-\eta_x)\sum_{i=m+1}^{m+\ell}\frac{\alpha_i}{\ell}}{\eta_{x^{\prime}}\sum_{j\in {x^{\prime}}}\frac{\beta_j}{|{x^{\prime}}|}+(1-\eta_{x^{\prime}})\sum_{j=m+1}^{m+\ell}\frac{\beta_j}{\ell}}
	\end{align}
	where $\eta_x=\frac{|x|}{\max\{|x|,\ell\}}$ and $\eta_{x^{\prime}}=\frac{|x^{\prime}|}{\max\{|x^{\prime}|,\ell\}}$
\end{lemma}
\begin{IEEEproof}
	Denote vector $\mathbf{v}_i=[0,\cdots,0,1,0,\cdots,0]$ with length $(m+\ell)$, where only the $i$-th position is 1  $(i\in\mathcal{I^{\prime}})$. From the Padding-and-Sampling protocol in Algorithm \ref{alg:PS},
	{\small
	\begin{align*}
	&\quad\Pr(y|x)=\sum\nolimits_{x_s\in x\cup \mathcal{S}}\Pr(x_s \text{ is sampled})\cdot\Pr(y|x_s) \\
	&=\eta_x\sum\nolimits_{i\in x}\frac{\Pr(y|x_s=i)}{|x|}+(1-\eta_x)\sum\nolimits_{i=1}^{\ell}\frac{\Pr(y|x_s=\perp_i)}{\ell}  \\
	&=\eta_x\sum\nolimits_{i\in x}\frac{\Pr(\mathbf{y}|\mathbf{v}_i)}{|x|}+(1-\eta_x)\sum\nolimits_{i=m+1}^{m+\ell}\frac{\Pr(\mathbf{y}|\mathbf{v}_i)}{\ell}
	\end{align*}
	}where $\eta_x$ is defined in Lemma \ref{lem:ratio}. 
	On the other hand,
	{\small
	\begin{align*}
	&\quad\Pr(\mathbf{y}|\mathbf{v}_i)
	=\Pr(\mathbf{y}[i]|\mathbf{x}[i]=1)\prod\nolimits_{k\in\mathcal{I^{\prime}}\backslash i}\Pr(\mathbf{y}[k]|\mathbf{x}[k]=0)\notag\\
	&=\frac{\Pr(\mathbf{y}[i]|\mathbf{x}[i]=1)}{\Pr(\mathbf{y}[i]|\mathbf{x}[i]=0)}\Phi
	=\frac{a_i^{\mathbf{y}[i]}(1-a_i)^{1-\mathbf{y}[i]}}{b_i^{\mathbf{y}[i]}(1-b_i)^{1-\mathbf{y}[i]}}\Phi\notag
	=\alpha_i^{\mathbf{y}[i]}\beta_i^{1-\mathbf{y}[i]}\Phi
	\end{align*}
	}where $\Phi\triangleq\prod\nolimits_{k\in\mathcal{I^{\prime}}}\Pr(\mathbf{y}[k]|\mathbf{x}[k]=0)> 0$, and $\alpha_i,\beta_i$ are defined in (\ref{equ:alpha_beta}). Since the value of $\mathbf{y}[k]$ is either 1 or 0 and $\alpha_i>\beta_i$, then $\beta_i\leqslant\frac{\Pr(\mathbf{y}|\mathbf{x}=\mathbf{v}_i)}{\Phi}\leqslant\alpha_i\quad(\forall i\in\mathcal{I^{\prime}})$. Thus, we have
	$\frac{\Pr(y|x)}{\Phi}\leqslant\eta_x\sum_{i\in x}\frac{\alpha_i}{|x|}+(1-\eta_x)\sum_{i=m+1}^{m+\ell}\frac{\alpha_i}{\ell}$ and $\frac{\Pr(y|x)}{\Phi}\geqslant\eta_x\sum_{i\in x}\frac{\beta_i}{|x|}+(1-\eta_x)\sum_{i=m+1}^{m+\ell}\frac{\beta_i}{\ell}$. Finally, 
	{\small
	\begin{align*}
	\frac{\Pr(y|x)}{\Pr(y|x^{\prime})}
	=\frac{\frac{\Pr(y|x)}{\Phi}}{\frac{\Pr(y|x^{\prime})}{\Phi}}
	\leqslant\frac{\eta_x\sum_{i\in x}\frac{\alpha_i}{|x|}+(1-\eta_x)\sum_{i=m+1}^{m+\ell}\frac{\alpha_i}{\ell}}{\eta_{x^{\prime}}\sum_{j\in {x^{\prime}}}\frac{\beta_j}{|{x^{\prime}}|}+(1-\eta_{x^{\prime}})\sum_{j=m+1}^{m+\ell}\frac{\beta_j}{\ell}}
	\end{align*}}
\end{IEEEproof}

Considering $\frac{\alpha_i}{\beta_j}=\frac{a_i(1-b_j)}{b_i(1-a_j)}$ is the upper bound of $\frac{\Pr(y|x=i)}{\Pr(y|x^{\prime}=j)}$, the distinguishability of a pair of item-set inputs $x$ and $x^{\prime}$ in (\ref{equ:ratio}) can be regarded as the combined distinguishability of the items that belong to the two sets. The parameter $\eta_x$ can be explained as the probability of sampling $i\in\mathcal{I}$ from the padded set $x_p$ of input $x$. If both $|x|$ and $|x^{\prime}|$ are greater or equal to $\ell$ (i.e., $\eta_x=\eta_{x^{\prime}}=1$), the distinguishability of the pair is averaged only among the items in the set; if not (then $\eta_x<1$ or $\eta_{x^{\prime}}<1$), the distinguishability of the dummy items will be involved since the original set would be padded with dummy items. From Lemma \ref{lem:ratio}, we observe that the distinguishability in IDUE-PS is determined by the privacy levels of the items in the pair of inputs (besides the number of items in the input set), which motivates that IDUE-PS satisfies the notion of MinID-LDP in some way (discussed below).

\textbf{Privacy Analysis.} In (\ref{equ:ratio}), the upper bound of the probability ratio $\frac{\Pr(y|x)}{\Pr(y|x^{\prime})}$ are related to the perturbation probabilities  of dummy items, i.e., $a_i$ and $b_i$ for $i=m+1,\cdots,m+\ell$. Since the dummy items themselves are not sensitive, we can select some reasonable values as their privacy levels. In this paper, we assume the privacy levels and perturbation probabilities of different dummy items are the same, denoted as $\epsilon_i=\epsilon^{*},~
a_i=a^{*},~b_i=b^{*}~(i=m+1,\cdots,m+\ell)$, then (\ref{equ:ratio}) can be rewritten as 
\begin{align}
\label{equ:ratio_new}
\frac{\Pr(y|x)}{\Pr(y|x^{\prime})}
\leqslant\frac{\eta_x\sum_{i\in x}\frac{\alpha_i}{|x|}+(1-\eta_x)\alpha^{*}}{\eta_{x^{\prime}}\sum_{j\in {x^{\prime}}}\frac{\beta_j}{|{x^{\prime}}|}+(1-\eta_{x^{\prime}})\beta^{*}}
\end{align}
where $\alpha^{*}=\frac{a^{*}}{b^{*}}$ and $\beta^{*}=\frac{1-a^{*}}{1-b^{*}}$. We consider the following expression of privacy budget for an item-set
\begin{align}
\label{equ:epsilon_x}
\epsilon_{x}=\ln\left[\eta_x\sum\nolimits_{i\in x}e^{\epsilon_i}/|x| + (1-\eta_x)e^{\epsilon^{*}}\right]
\quad(\forall x\in\mathcal{D})
\end{align}
which can be regarded as the combined privacy budget of the items in the set $x$ (the privacy budget of dummy items will be involved when $|x|<\ell$, i.e., $\eta_x<1$). The combined privacy budget in  (\ref{equ:epsilon_x}) is larger than the averaged privacy budget $\sum_{i\in x}\epsilon_i/|x|$ because  the exponential function $f(\epsilon)=e^{\epsilon}$ is convex with property $\sum_{i}k_if(\epsilon_i)\geqslant f(\sum_{i}k_i\epsilon_i)$, where $0\leqslant k_i\leqslant1$ and $\sum_{i}k_i=1$. Based on the results in Lemma \ref{lem:ratio}, we show the fact that IDUE-PS satisfies MinID-LDP.

\begin{theorem} 
\label{thm:MinID-LDP}
If mechanism IDUE  with perturbation probabilities $a_i,b_i~(i\in\mathcal{I})$ satisfies MinID-LDP for single-item input with privacy budget $\epsilon_1,\epsilon_2,\cdots,\epsilon_m$, i.e.,
\begin{align}
\label{equ:constraint_single}
\frac{\alpha_i}{\beta_j}
=\frac{a_i(1-b_j)}{b_i(1-b_j)}
\leqslant e^{\min\{\epsilon_i,\epsilon_j\}}\quad(\forall i,j\in\mathcal{I})
\end{align}
then IDUE-PS with the same perturbation probabilities will satisfy MinID-LDP for item-set input, i.e.,
\begin{align}
\label{equ:constraint_set}
\frac{\Pr(y|x)}{\Pr(y|x^{\prime})}\leqslant e^{\min\{\epsilon_x,\epsilon_{x^{\prime}}\}}\quad(\forall x,x^{\prime}\in\mathcal{D}, \forall y)
\end{align} 
where privacy budget of item-set is defined in (\ref{equ:epsilon_x}) and the privacy budget of dummy items $\epsilon^{*}\in\{\epsilon_1,\epsilon_2,\cdots,\epsilon_m\}$,
\end{theorem}
\begin{IEEEproof}
	Denote $\alpha_{\text{max}}=\max_{i\in\mathcal{I}}\{\alpha_i\},\beta_{\text{min}}=\min_{j\in\mathcal{I}}\{\beta_j\}$. According to $\alpha_i/\beta_j\leqslant e^{\min\{\epsilon_i,\epsilon_j\}}$, we have $\alpha^{*}/\beta_{\text{min}}\leqslant e^{\epsilon^{*}}$ and $\alpha_i/\beta_{\text{min}}\leqslant e^{\epsilon_i}~(\forall i\in\mathcal{I})$. Then (\ref{equ:ratio_new}) can be rewritten as
	{\small
	\begin{align*}
	&\quad\frac{\Pr(y|x)}{\Pr(y|x^{\prime})}
	\leqslant\frac{\eta_x\sum_{i\in x}\frac{\alpha_i}{|x|}/\beta_{\text{min}}+(1-\eta_x)\alpha^{*}/\beta_{\text{min}}}{\eta_{x^{\prime}}\sum_{j\in {x^{\prime}}}\frac{\beta_j}{|{x^{\prime}}|}/\beta_{\text{min}}+(1-\eta_{x^{\prime}})\beta^{*}/\beta_{\text{min}}}\\
	&\leqslant\frac{\eta_x\sum\limits_{i\in x}\frac{e^{\epsilon_i}}{|x|}+(1-\eta_x)e^{\epsilon^{*}}}{\eta_{x^{\prime}}\sum\limits_{j\in {x^{\prime}}}\frac{1}{|{x^{\prime}}|}+(1-\eta_{x^{\prime}})}
	=\frac{\eta_x\sum\limits_{i\in x}\frac{e^{\epsilon_i}}{|x|}+(1-\eta_x)e^{\epsilon^{*}}}{1}
	= e^{\epsilon_{x}}
	\end{align*}
	}where $\epsilon_{x}$ is defined in (\ref{equ:epsilon_x}). On the other hand, according to $\alpha_i/\beta_j\leqslant e^{\min\{\epsilon_i,\epsilon_j\}}$, we have  $\beta^{*}/\alpha_{\text{max}}\geqslant e^{-\epsilon^{*}}$ and $\beta_j/\alpha_{\text{max}}\geqslant e^{-\epsilon_j}~(\forall j\in\mathcal{I})$. Then (\ref{equ:ratio_new}) can be rewritten as
	{\small
	\begin{align*}
	&\quad\frac{\Pr(y|x)}{\Pr(y|x^{\prime})}
	\leqslant\frac{\eta_x\sum_{i\in x}\frac{\alpha_i}{|x|}/\alpha_{\text{max}}+(1-\eta_x)\alpha^{*}/\alpha_{\text{max}}}{\eta_{x^{\prime}}\sum_{j\in {x^{\prime}}}\frac{\beta_j}{|{x^{\prime}}|}/\alpha_{\text{max}}+(1-\eta_{x^{\prime}})\beta^{*}/\alpha_{\text{max}}}\\
	&\leqslant\frac{\eta_x\sum_{i\in x}\frac{1}{|x|}+(1-\eta_x)\cdot 1}{\eta_{x^{\prime}}\sum\limits_{j\in {x^{\prime}}}\frac{e^{-\epsilon_j}}{|{x^{\prime}}|}+(1-\eta_{x^{\prime}})e^{-\epsilon^{*}}}
	=\frac{1}{\eta_{x^{\prime}}\sum\limits_{j\in {x^{\prime}}}\frac{e^{-\epsilon_j}}{|{x^{\prime}}|}+(1-\eta_{x^{\prime}})e^{-\epsilon^{*}}}\\
	&\leqslant\eta_{x^{\prime}}\sum\nolimits_{j\in {x^{\prime}}}\frac{e^{-\epsilon_j}}{|{x^{\prime}}|}+(1-\eta_{x^{\prime}})e^{-\epsilon^{*}}
	= e^{\epsilon_{x^{\prime}}}
	\end{align*}
	}The last inequality is obtained by Cauchy-Schwarz inequality. Finally, $\frac{\Pr(y|x)}{\Pr(y|x^{\prime})}\leqslant\min\{e^{\epsilon_{x}},e^{\epsilon_{x^{\prime}}}\}=e^{\min\{\epsilon_{x},\epsilon_{x^{\prime}}\}}$.
\end{IEEEproof}

According to Theorem \ref{thm:MinID-LDP}, the perturbation probabilities in IDUE-PS for item-set input can be determined with the same way in IDUE, i.e., solving the optimization problems  (\ref{equ:opt_ab}) with only $2t$ variables and $t^2$ constraints to get the optimal solution ($t$ is the number of privacy levels), or the constrained models (\ref{equ:opt_tau}) and (\ref{equ:opt_b}) with less computational cost to get the near-optimal solution. For the privacy budget of dummy items, theoretically, we can select $\epsilon^{*}$ to be any value from $\{\epsilon_1,\epsilon_2,\cdots,\epsilon_m\}$.  Though a larger $\epsilon^{*}$ will improve the utility of \emph{dummy} items, the result of frequency estimation for dummy items will be ignored in aggregation because they are not our task. Also, the value of  $\epsilon^{*}$ (selected from the original budgets) does not change the optimization problem and the optimal solution because the objective function (only depends on original items) and  constraints (only depends on privacy levels) are the same. Therefore, we select $\epsilon^{*}=\min\{\epsilon_1,\epsilon_2,\cdots,\epsilon_m\}$ to guarantee the privacy with smaller budget $\epsilon_{x}$ in (\ref{equ:epsilon_x}).

\section{Evaluation}

In this section, we evaluate the performance of frequency estimation of IDUE and compare it with RAPPOR \cite{erlingsson2014rappor} and OUE \cite{wang2017locally}. Note that RAPPOR and OUE satisfy $\epsilon$-LDP with $\epsilon=\min\{\mathcal{E}\}$, while IDUE and IDUE-PS satisfy $\mathcal{E}$-MinID-LDP. The perturbation probabilities in IDUE (and IDUE-PS) can be obtained by three optimization models in  (\ref{equ:opt_ab}),  (\ref{equ:opt_tau}), and  (\ref{equ:opt_b}), denoted by \verb|opt0|, \verb|opt1| and \verb|opt2| respectively.

\textbf{Applicability of Multiple Privacy Budgets.} 
Though our notion MinID-LDP generally considers different privacy budgets $\epsilon$ for different items, in practice, these items can be classified by a small number of categories with distinct privacy levels. For example, thousands of clinical conditions can be classified by three  categories including serious diseases, moderate diseases and common symptoms, where only three values of privacy budgets  need to be determined according to the applications. We note that the privacy benefit is bounded by  $2\min\{\mathcal{E}\}$ even when other privacy budgets are higher than $2\min\{\mathcal{E}\}$ (refer to Lemma \ref{lem:relationship}). In the case of item-set, we consider the privacy budget of a set of items with the form of \eqref{equ:epsilon_x}, which is a combination of privacy budgets of items in this set. Theorem \ref{thm:MinID-LDP} shows that the perturbation probabilities of IDUE-PS that satisfies MinID-LDP can be determined by IDUE (where items classified in the same privacy level have the same perturbation probabilities). Therefore, the complexity of our solution, including the number of assigned privacy budgets and computation cost of solving our model, only depends on  the number of privacy levels (rather than the domain size of single-item or item-set).

\textbf{Datasets.} 
We conduct the experiments over two synthetic single-item datasets (with different distributions and domain sizes) and three real item-set datasets (obtained from public data sources), whose parameters are shown in Table \ref{tab:data}. The data with Power-law distribution is obtained by generating random values from the power-law distribution with the law's exponent $\alpha=2$, then scaling and rounding into an integer that belongs to  $\mathcal{I}=\{1,2,\cdots,m\}$. The data with Uniform distribution of each user is uniformly generated from $\mathcal{I}=\{1,2,\cdots,m\}$.






\begin{table}[t!]
    \small
    \centering
    \caption{Synthetic and Real-world Datasets}
    \vspace{-2mm}
    \begin{tabular}{c|ccc}
    \hline
    \textbf{Datasets} & \textbf{\# Records} & \textbf{\# Users ($n$)} &  \textbf{\# Items ($m$)} \\
    \hline
    Power-law & 100,000 & 100,000 & 100 \\
    Uniform & 100,000 & 100,000 & 1,000 \\ 
    Retail \cite{itemset_data}  & 908,576  & 88,162 & 16,470   \\
    Kosarak \cite{itemset_data} & 8,019,015 & 990,002  & 41,270 \\
    Clothing \cite{clothing}  & 192,544  & 105,508  &  5,850  \\
    \hline
    \end{tabular}
    \label{tab:data}
\end{table}

\textbf{Evaluation Metrics.} 
We use the \emph{total} Mean Squared Error (MSE) of all items and the \emph{average}  Relative Error (RE) of top $k$ frequent items, defined by
\begin{align*}
    \text{MSE} = \sum\nolimits_{i\in\mathcal{I}}\frac{(\hat{c}_i-c^{*}_i)^2}{n},\quad
    \text{RE}(k) = \frac{1}{k}\sum\nolimits_{i\in\mathcal{T}(k)}\frac{|\hat{c}_i-c^{*}_i|}{c^{*}_i}
\end{align*}
where $\hat{c}_i$ (resp. $c^{*}_i$) is the estimated (resp. true) count of item $i$, and $\mathcal{T}(k)$ is the set of ground true top $k$ frequent items. We also use the ranking of estimated frequencies to identify top $k$ frequent items and evaluate its \emph{precision} (in Sec. \ref{sec:itemset_data}). All experimental results are averaged with ten repeats.

\textbf{Setting of Privacy Budget.} 
We consider multiple privacy levels of the inputs, thus we  need to assign multiple privacy budgets to them. Assume there are three privacy levels with privacy budget $\{\epsilon,1.2\epsilon,2\epsilon\}$ (as default values), where $\epsilon$ is the smallest privacy budget. The privacy budget for all items are randomly selected from the three values with a certain budget distribution, where the default distribution is $\{5\%,5\%,90\%\}$, and we will change the budget distribution in the experiments to evaluate the impact.


\subsection{Single-item Data}

\begin{figure}[!t]
    \centering
	\includegraphics[width=0.49\textwidth]{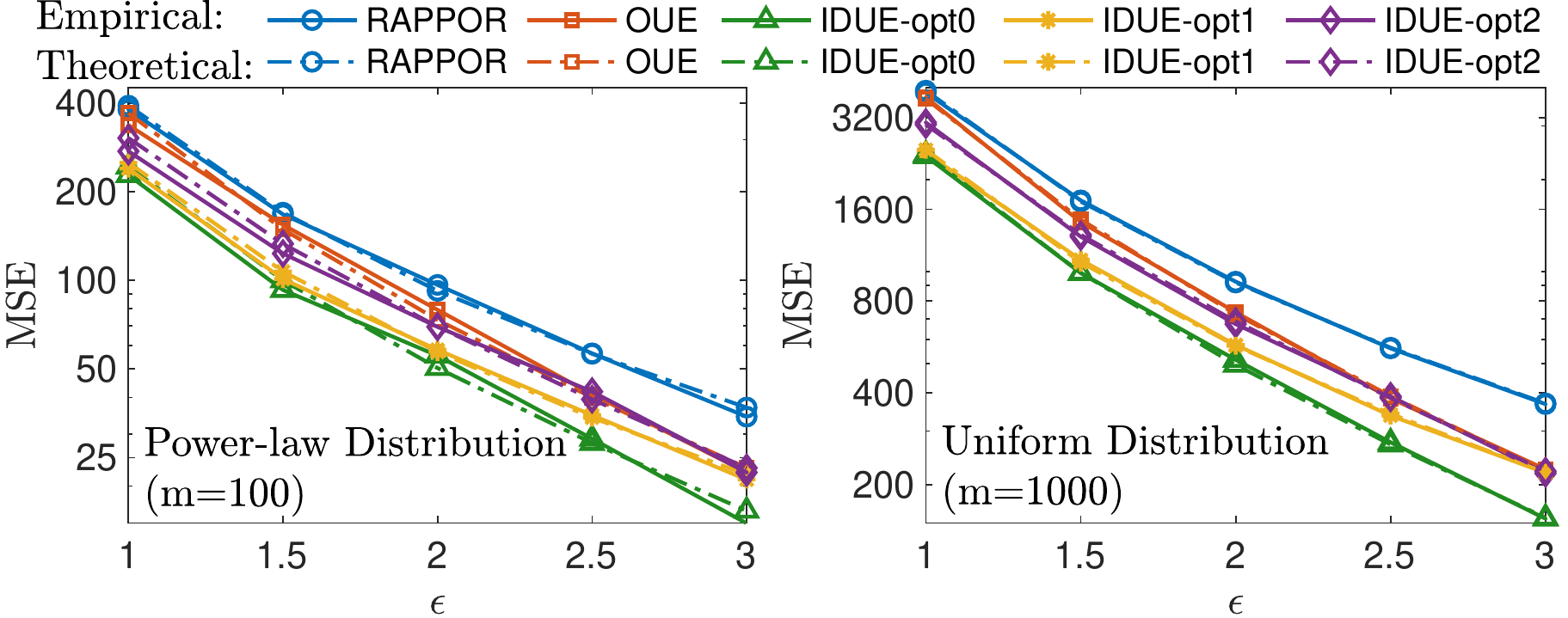}
	\\ \vspace{-1mm}
	\caption{Comparison of Empirical (dashed lines) and Theoretical (solid lines) results of synthetic  data (single-item input).}
	\label{fig:single1}
	\vspace{-3mm}
\end{figure}

\begin{figure}[!t]
    \centering
	\includegraphics[width=0.49\textwidth]{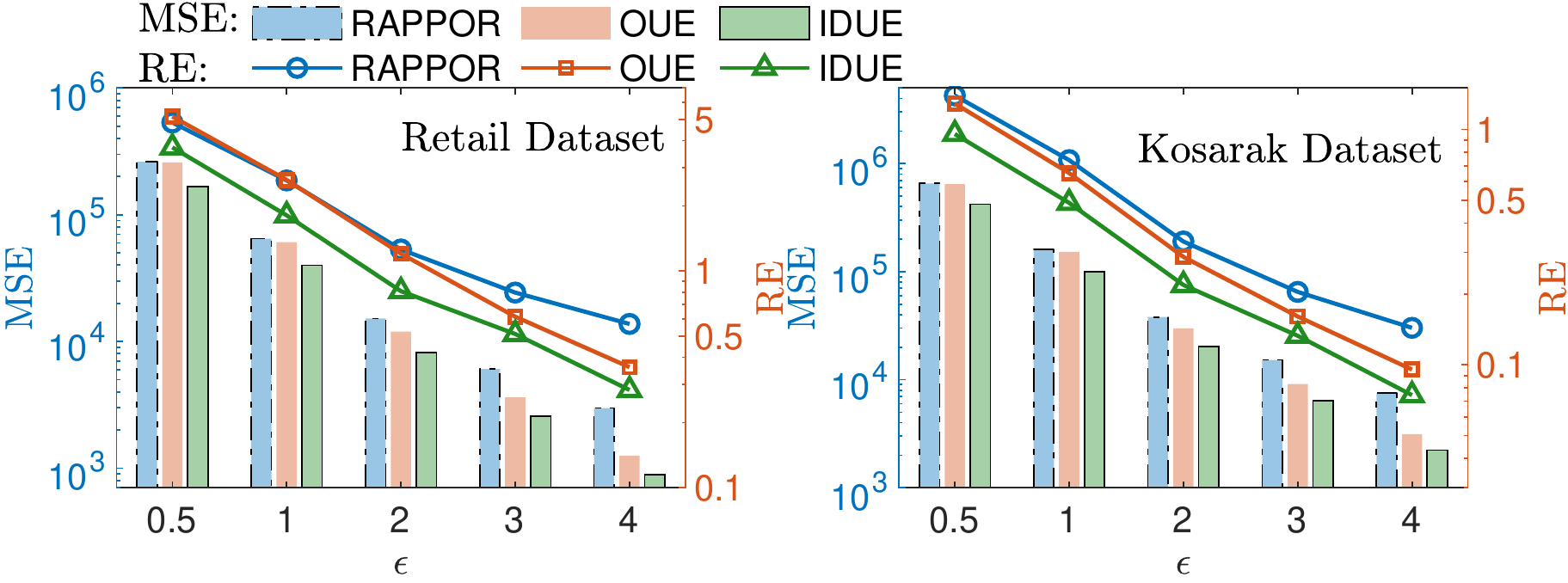}
	\\ \vspace{-1mm}
	\caption{MSE and RE of real-world datasets (single-item input).}
	\label{fig:single2_MSE}
	\vspace{-3mm}
\end{figure}

\textbf{Validation of Theoretical Analysis.} 
Fig. \ref{fig:single1} shows the empirical  and theoretical results of the MSE of the estimated frequency under Power-law and Uniform distributions. The empirical results (solid lines) are very close to the theoretical results (dashed lines), which validates the correctness of our theoretical analysis.  We can observe that mechanisms satisfying LDP and MinID-LDP have relatively similar utility but IDUE with MinID-LDP outperforms RAPPOR and OUE by adjusting the perturbation probabilities for different inputs. For IDUE, the reduced optimization models (i.e., \verb|opt1| and \verb|opt2|) have relatively larger MSEs than the original optimization model (i.e., \verb|opt0|) due to the further constrained variable space, but they still can provide the near-optimal solution for IDUE with less computational complexity. In the following experiments, we only evaluate IDUE sovled by \verb|opt0| for simplicity of plots.

\begin{figure}[!t]
    \centering
	\includegraphics[width=0.49\textwidth]{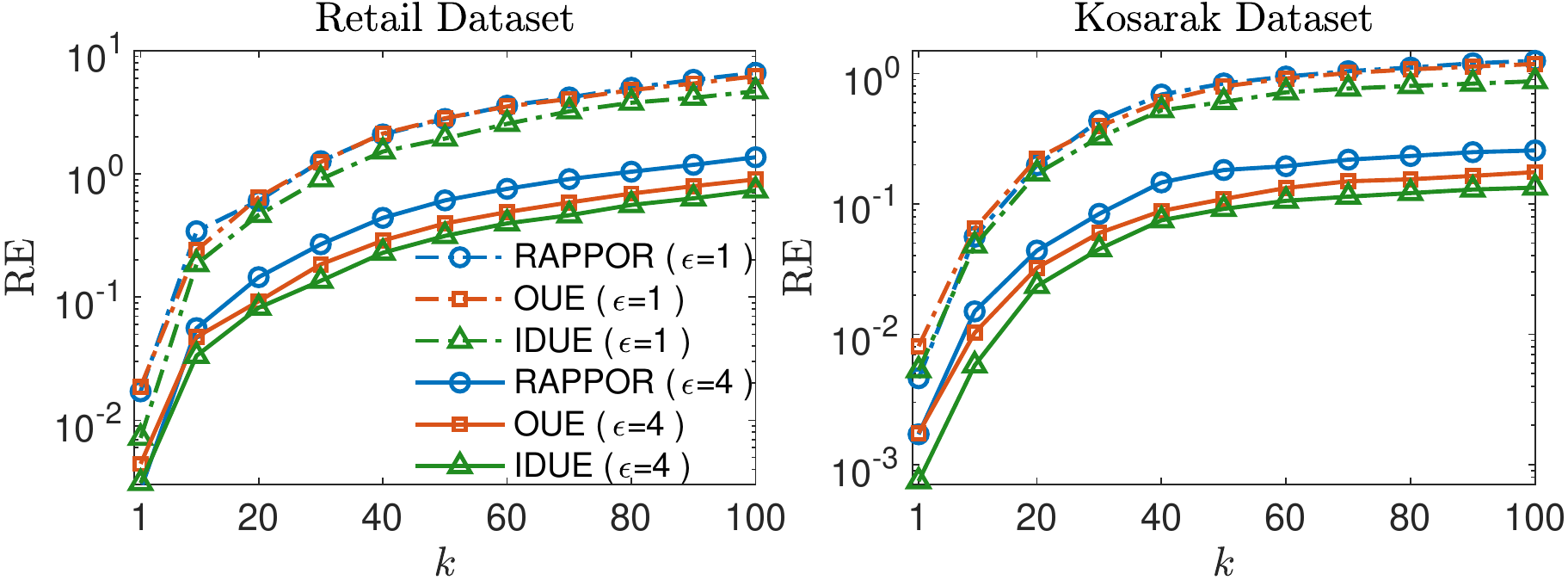}
	\\ \vspace{-1mm}
	\caption{RE of top $k$  frequent items (varying $k$).}
	\label{fig:single2_RE}
\end{figure}

\textbf{Results on Real-world Datasets.}
Fig. \ref{fig:single2_MSE} shows the total MSE of all items and average RE of top $k$ frequent items (with $k=20$) of Retail and Kosarak datasets, where only the first item of each user is considered in the case that each input is a single-item. We also show the results of RE under different $k$ in Fig. \ref{fig:single2_RE}. The proposed IDUE has the best utility (i.e., smallest MSE and RE of frequency estimation)  for all considered $\epsilon$ and $k$.

\begin{figure}[!t]
    \centering
	\includegraphics[width=0.485\textwidth]{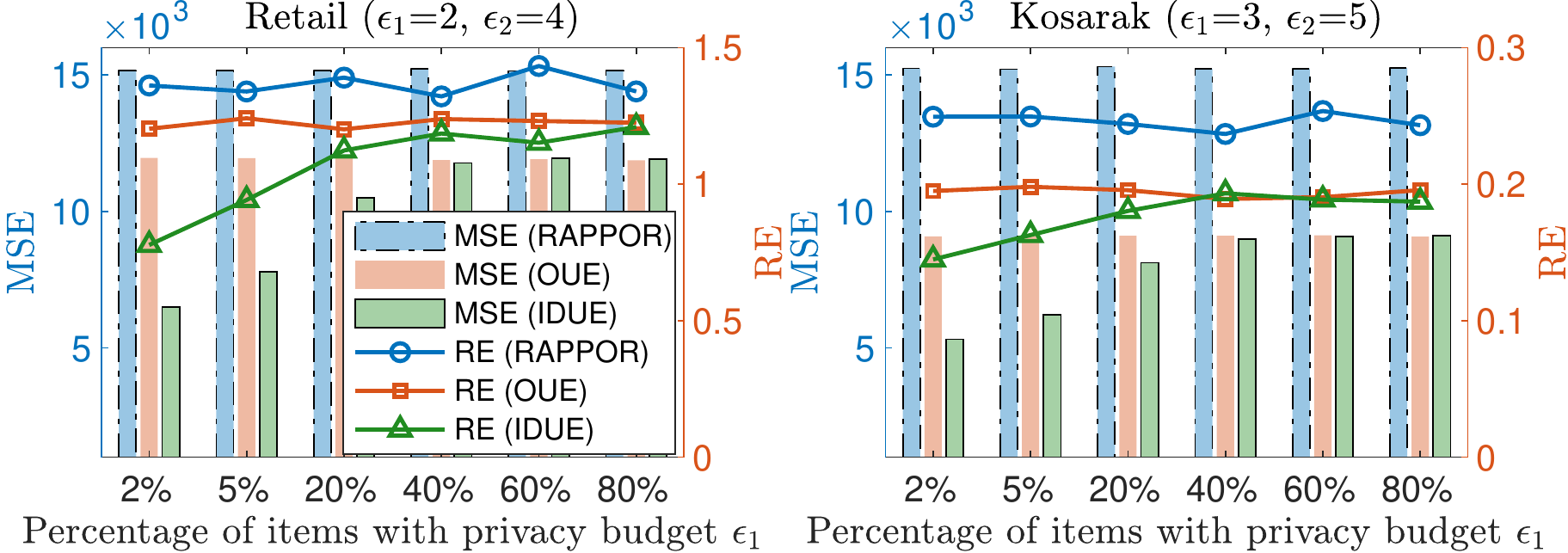}
	\\ \vspace{-1mm}
	\caption{Under different privacy budget distributions.}
	\label{fig:single3}
\end{figure}

\textbf{Influence of Privacy Budget Distributions.} 
The MSE and RE (with $k=50$) under different budget distributions are shown in Fig. \ref{fig:single3}, where we only consider two privacy levels (with privacy budgets $\epsilon_1<\epsilon_2$) and vary the percentage of items whose privacy budget is $\epsilon_1$ (the smaller one). Under a smaller percentage, i.e., only a few of items are more sensitive than others, IDUE can get more benefits from our relaxed privacy notion MinID-LDP. However, when the percentage of the more sensitive items is large (such as more than $40\%$), IDUE has almost the same total MSE and average RE as OUE (which outperforms RAPPOR). 

\subsection{Item-set Data}
\label{sec:itemset_data}

\begin{figure}[!t]
    \centering
	\includegraphics[width=0.485\textwidth]{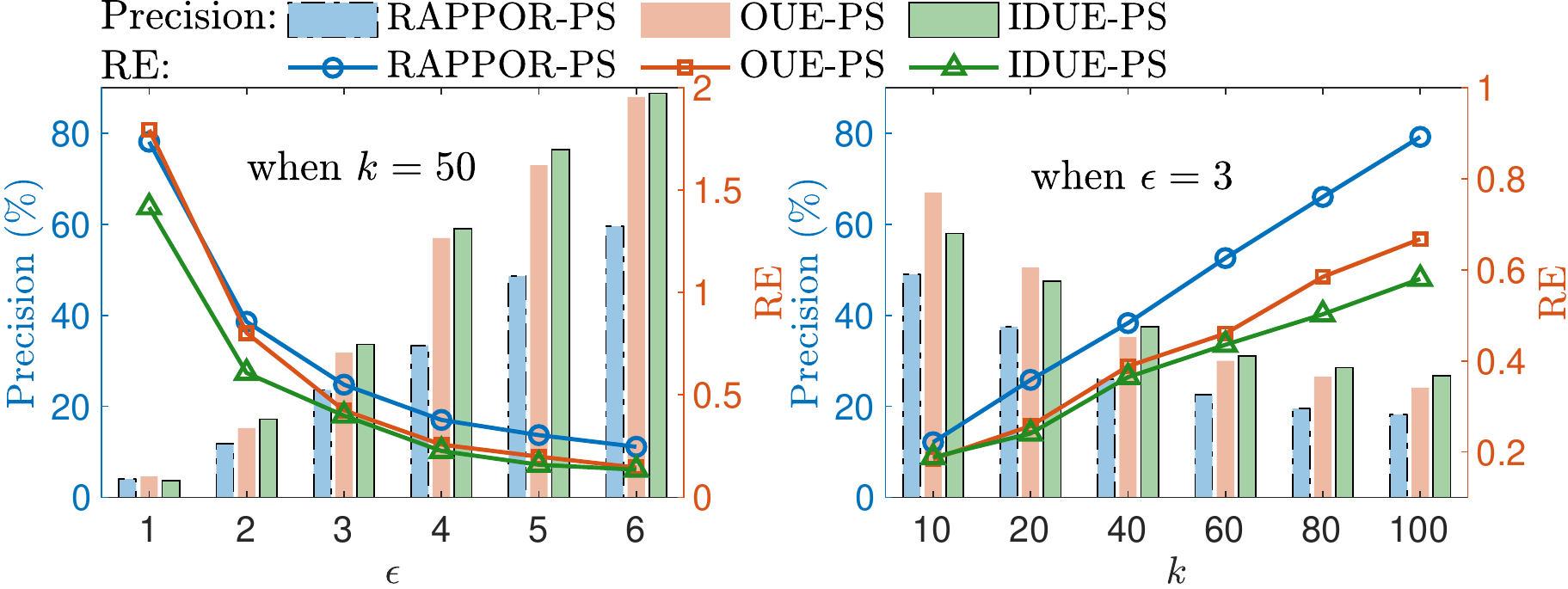}
	\\ \vspace{-1mm}
	\caption{Precision and RE of Clothing itemset data ($\ell=2$).}
	\label{fig:set1_clothing}
\end{figure}


\textbf{Accuracy of Top Frequent Items Identification.} Fig. \ref{fig:set1_clothing} shows the precision of top $k$ frequent items identification (i.e., the proportion of correct selections over all predicted top frequent items, obtained from the ranking of estimated frequencies) and RE (with the above $k$) of Clothing dataset under different $\epsilon$ and $k$. We note that each item has privacy budget in $\{\epsilon,1.2\epsilon,2\epsilon\}$ with distribution $\{5\%,5\%,90\%\}$ (the default one), where $\epsilon$ is the budget for a single item, and the privacy budget of each item-set is a combination of items' budgets in the set, defined in \eqref{equ:epsilon_x}. The proposed mechanism IDUE-PS has the smallest RE (similar to the previous results) among three mechanisms. But for a smaller $\epsilon$ or $k$ (such as $\epsilon=1$ in the left plot and $k=10$ or $20$ in the right plot), the precision of top frequent items identification may be worse than the precision of OUE-PS (noth that for larger $\epsilon$ and $k$, IDUE-PS has the highest precision). Such an observation might be caused by the distinct protection for different items in IDUE-PS, where items with the smallest privacy budget have larger error than the other two mechanisms (which was explained in Sec. \ref{sec:comparison_mechanisms}). But when $\epsilon$ or $k$ is larger, such impact  will be mitigated (compared with other mechanisms) because a larger $\epsilon$ allows less noise to be added in the perturbation of items with the smallest privacy budget. On the other hand, a larger $k$ generally leads to a lower precision on top frequent items identification because many items in real-world data have the middle ranking, thus the same amount of error on estimated frequencies will make a big difference on the estimated ranking. However, a larger $k$ also makes top items with the smallest privacy budget have more chances to be selected, thus IDUE-PS can get  benefits from the balance of distinct amount of noise of different items (caused by distinct privacy protection levels).

\begin{figure}[!t]
    \centering
	\includegraphics[width=0.485\textwidth]{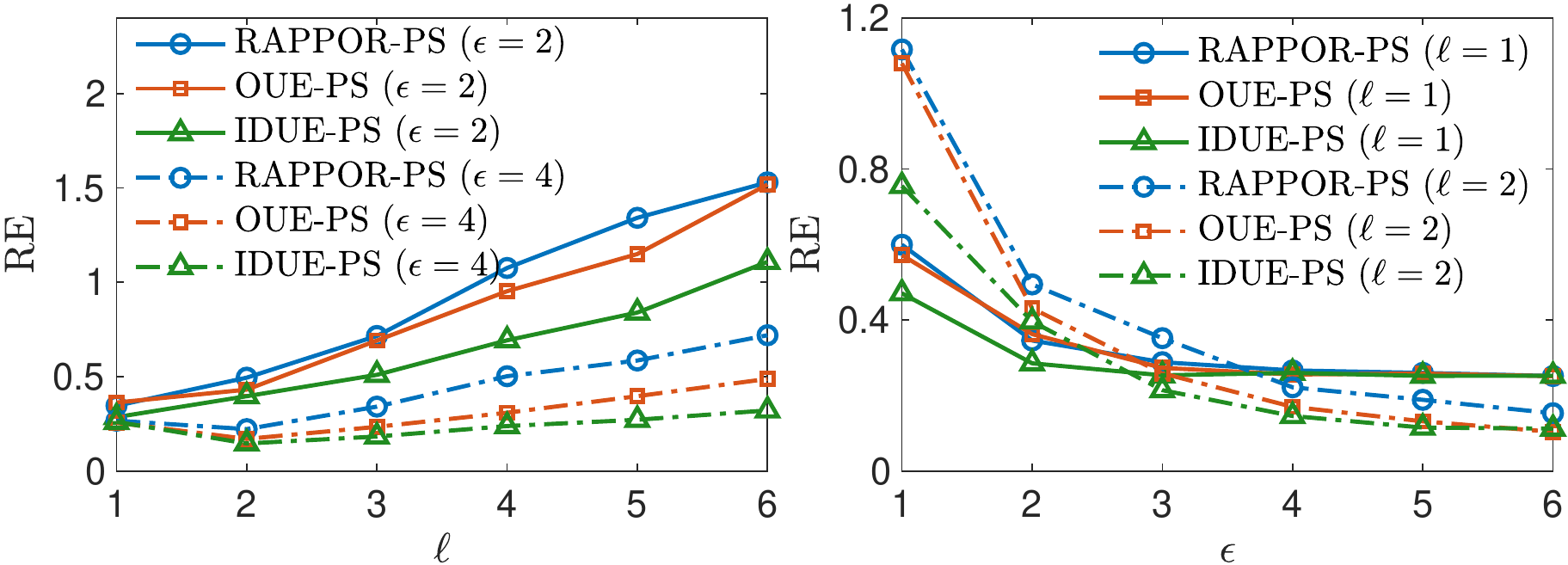}
	\\ \vspace{-1mm}
	\caption{Varying $\ell$ and $\epsilon$ in Clothing itemset data ($k=20$).}
	\label{fig:set2_clothing}
	\vspace{-1mm}
\end{figure}

\begin{figure}[!t]
    \centering
	\includegraphics[width=0.485\textwidth]{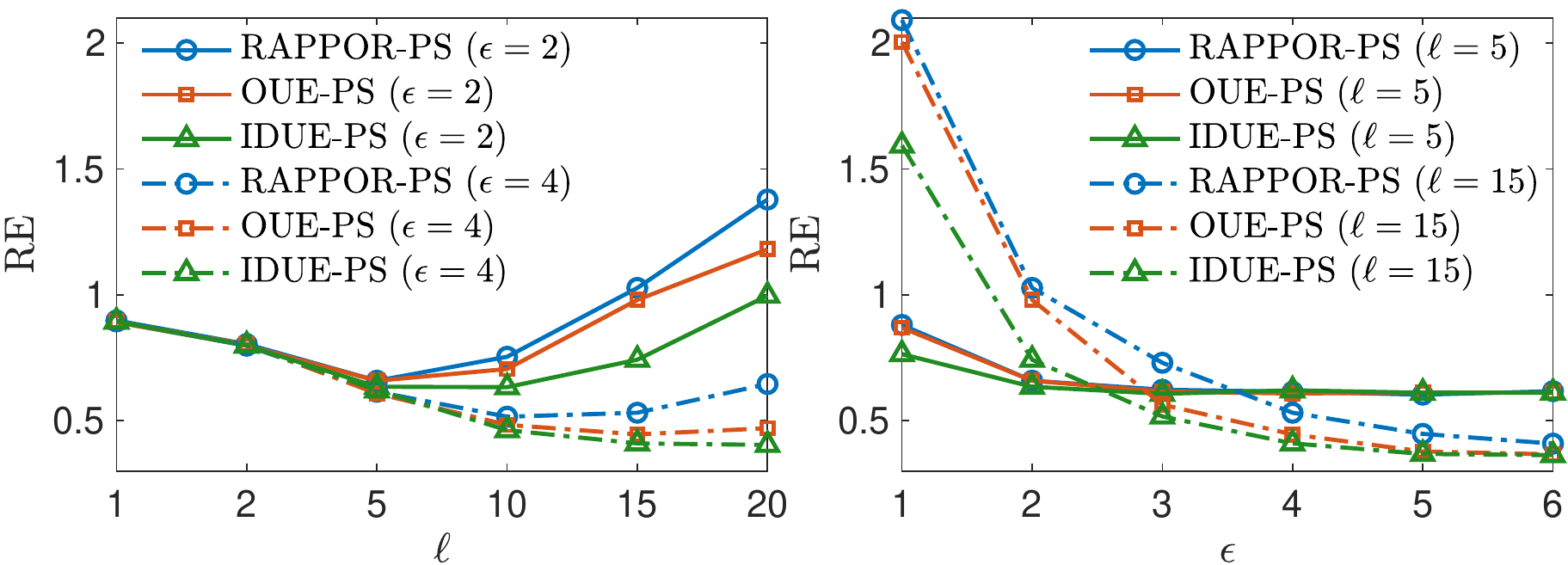}
	\\ \vspace{-1mm}
	\caption{Varying $\ell$ and $\epsilon$ in Kosarak itemset data ($k=100$).}
	\label{fig:set2_kosarak}
\end{figure}

\textbf{Influence of the Padding Length $\ell$ for Item-Set Data.} The results of Clothing and Kosarak datasets, where each user approximately has 2 and 8 items in average respectively, under different padding length $\ell$ are shown in Fig. \ref{fig:set2_clothing} and Fig. \ref{fig:set2_kosarak} (results in Retail data have similar trends as in Kasarak data). We can observe that the optimal or near-optimal $\ell$ differs in both data distribution (users in Kosarak dataset have more items than in Clothing dataset) and privacy budget $\epsilon$ (the optimal $\ell$ is larger under a larger $\epsilon$). The second observation is caused by the influence of the given $\epsilon$ on the trade-off between variance and bias of frequency estimation (i.e., a larger $\ell$ leads to a larger variance while a smaller $\ell$ leads to a larger bias \cite{wang2018locally}). Under a smaller $\epsilon$ (i.e., stronger privacy), the error from variance dominates the error from bias, thus a smaller $\ell$ should be selected to reduce the variance. Similarly, under a larger $\epsilon$, a larger $\ell$ should be selected to reduce the bias. Also, when fixing a relatively small $\ell$ (such as $\ell=5$ in Kosarak), RE does not reduce much with increasing $\epsilon$ because $\epsilon$ has little influence on the bias (which largely contributes to the error in this case). In \cite{wang2018locally}, $\ell$ is selected as the 90th percentile of numbers of items of all users (i.e., only depending on data distribution). However, a good $\ell$ should also depend on $\epsilon$ from above discussions. A simple empirical strategy is to select $\ell$ as the average number of items in each user under a larger $\epsilon$ (such as 4), while select $\ell$ less than the average one under a smaller $\epsilon$ (such as 2). The advanced strategy of how to determine the optimal $\ell$ under a specific $\epsilon$ will be our future work.


\section{Discussions}

\textbf{Additional Gain from Incomplete Privacy Policy Graph.}
According to Lemma \ref{lem:relationship}, the gain of MinID-LDP compared with LDP is at most twice of the privacy budget, which is caused by the required privacy protection on all pairs of inputs (i.e., complete graph shown in Fig. \ref{fig:definitions}). However, if some of the pairs do not need to be protected (such incomplete graph can be defined by the secret policy in Blowfish privacy \cite{he2014blowfish}), the gain of MinID-LDP can be larger than $2\min\{\mathcal{E}\}$ because some inputs might not need to be indistinguishable from the inputs with the smallest privacy budget.

\textbf{Other Instantiations of ID-LDP.}
Besides MinID-LDP, other instantiations of ID-LDP can be defined. For example, we can define AvgID-LDP as ID-LDP with the average function, i.e., $r(\epsilon_x,\epsilon_{x^{\prime}})=(\epsilon_x+\epsilon_{x^{\prime}})/2$, which bounds the privacy budget of a pair of inputs by the averaged budget of the two inputs. Similar to MinID-LDP, the notion of AvgID-LDP satisfies sequential composition like Theorem \ref{thm:composition_min}. Moreover, the perturbation mechanisms developed in Sec. \ref{sec:mechanism_single} and Sec. \ref{sec:mechanism_set} are also applicable to AvgID-LDP.

\textbf{Benefits of Our Framework.}
The utility improvement of IDUE is dependent on the utility metrics and the distributions of privacy budget and data. In the case of two different privacy budgets, if items with the smaller budget only have little influence on the utility (generally the number of these items is very small in this case), the utility of IDUE will approach the LDP mechanism with the larger budget.  Note that larger noise will be added in the perturbation of the items with the smaller budget to satisfy the privacy constraint, but the impact on utility is very small in this case.

\textbf{Limitations of Our Framework.}
First, the amount of benefits of our framework depends on budget distribution.  If majority of items have the smallest budget, the benefit obtained from  IDUE might be very small (see Fig. \ref{fig:single3}) because these items greatly affect the utility. Second, the distinct amount of noise for different items may have negative influence on the utility of some applications, such as the precision of top frequent item identification in Fig. \ref{fig:set1_clothing}.

\section{Conclusion}
In this paper, a new privacy notion named ID-LDP with an instantiation MinID-LDP is proposed to provide input-discriminative protection in the local setting. MinID-LDP is shown to satisfy the sequential composition theorem as LDP and can be regarded as the fine-grained version of LDP. We propose the perturbation mechanism framework IDUE that satisfies ID-LDP, where the perturbation probabilities are solved by the optimization problem with reasonable scale. Then, based on Padding-and-Sampling protocol, the mechanism is extended to  apply to item-set input, named IDUE-PS, to solve the scalability and utility problem for the item-set type of input. IDUE-PS is also shown to satisfy MinID-LDP. Finally,  experimental results validate the advantage of our privacy notion and mechanisms, compared with the existing ones. 

For future work, we will extend our work to handle more complex data types or analysis tasks.

\bibliographystyle{IEEEtran}
\bibliography{IEEEabrv,mybibfile}

\begin{thebibliography}{10}
\providecommand{\url}[1]{#1}
\csname url@samestyle\endcsname
\providecommand{\newblock}{\relax}
\providecommand{\bibinfo}[2]{#2}
\providecommand{\BIBentrySTDinterwordspacing}{\spaceskip=0pt\relax}
\providecommand{\BIBentryALTinterwordstretchfactor}{4}
\providecommand{\BIBentryALTinterwordspacing}{\spaceskip=\fontdimen2\font plus
\BIBentryALTinterwordstretchfactor\fontdimen3\font minus
  \fontdimen4\font\relax}
\providecommand{\BIBforeignlanguage}[2]{{%
\expandafter\ifx\csname l@#1\endcsname\relax
\typeout{** WARNING: IEEEtran.bst: No hyphenation pattern has been}%
\typeout{** loaded for the language `#1'. Using the pattern for}%
\typeout{** the default language instead.}%
\else
\language=\csname l@#1\endcsname
\fi
#2}}
\providecommand{\BIBdecl}{\relax}
\BIBdecl

\bibitem{dwork2006differential}
C.~Dwork, ``Differential privacy,'' in \emph{ICALP}, 2006, pp. 1--12.

\bibitem{dwork2006calibrating}
C.~Dwork, F.~McSherry, K.~Nissim, and A.~Smith, ``Calibrating noise to
  sensitivity in private data analysis,'' in \emph{Theory of Cryptography
  Conference (TCC)}, 2006, pp. 265--284.

\bibitem{chen2016private}
R.~Chen, H.~Li, A.~Qin, S.~P. Kasiviswanathan, and H.~Jin, ``Private spatial
  data aggregation in the local setting,'' in \emph{IEEE ICDE}, 2016, pp.
  289--300.

\bibitem{erlingsson2014rappor}
{\'U}.~Erlingsson, V.~Pihur, and A.~Korolova, ``Rappor: Randomized aggregatable
  privacy-preserving ordinal response,'' in \emph{ACM CCS}, 2014, pp.
  1054--1067.

\bibitem{apple2017learning}
``Learning with privacy at scale,''
  \url{https://machinelearning.apple.com/2017/12/06/learning-with-privacy-at-scale.html},
  2017.

\bibitem{wang2017locally}
T.~Wang, J.~Blocki, N.~Li, and S.~Jha, ``Locally differentially private
  protocols for frequency estimation,'' in \emph{USENIX Security Symposium},
  2017, pp. 729--745.

\bibitem{wang2018locally}
T.~Wang, N.~Li, and S.~Jha, ``Locally differentially private frequent itemset
  mining,'' in \emph{IEEE S\&P}, 2018, pp. 127--143.

\bibitem{ye2019privkv}
Q.~Ye, H.~Hu, X.~Meng, and H.~Zheng, ``Privkv: Key-value data collection with
  local differential privacy,'' in \emph{IEEE S\&P}, 2019.

\bibitem{wang2015personalized}
S.~Wang, L.~Huang, M.~Tian, W.~Yang, H.~Xu, and H.~Guo, ``Personalized
  privacy-preserving data aggregation for histogram estimation,'' in \emph{IEEE
  GLOBECOM}, 2015, pp. 1--6.

\bibitem{andres2013geo}
M.~Andr{\'e}s, N.~Bordenabe, K.~Chatzikokolakis, and C.~Palamidessi,
  ``Geo-indistinguishability: Differential privacy for location-based
  systems,'' in \emph{ACM CCS}, 2013, pp. 901--914.

\bibitem{chatzikokolakis2013broadening}
K.~Chatzikokolakis, M.~E. Andr{\'e}s, N.~E. Bordenabe, and C.~Palamidessi,
  ``Broadening the scope of differential privacy using metrics,'' in
  \emph{International Symposium on Privacy Enhancing Technologies Symposium},
  2013, pp. 82--102.

\bibitem{kifer2012rigorous}
D.~Kifer and A.~Machanavajjhala, ``A rigorous and customizable framework for
  privacy,'' in \emph{ACM SIGMOD-SIGACT-SIGAI symposium on Principles of
  Database Systems}, 2012, pp. 77--88.

\bibitem{he2014blowfish}
X.~He, A.~Machanavajjhala, and B.~Ding, ``Blowfish privacy: Tuning
  privacy-utility trade-offs using policies,'' in \emph{ACM SIGMOD}, 2014, pp.
  1447--1458.

\bibitem{bun2016concentrated}
M.~Bun and T.~Steinke, ``Concentrated differential privacy: Simplifications,
  extensions, and lower bounds,'' in \emph{Theory of Cryptography Conference},
  2016, pp. 635--658.

\bibitem{jorgensen2015conservative}
Z.~Jorgensen, T.~Yu, and G.~Cormode, ``Conservative or liberal? personalized
  differential privacy,'' in \emph{IEEE ICDE}, 2015, pp. 1023--1034.

\bibitem{duchi2013local}
J.~C. Duchi, M.~I. Jordan, and M.~J. Wainwright, ``Local privacy and
  statistical minimax rates,'' in \emph{IEEE FOCS}, 2013, pp. 429--438.

\bibitem{gu2019pckv}
X.~Gu, M.~Li, Y.~Cheng, L.~Xiong, and Y.~Cao, ``Pckv: Locally differentially
  private correlated key-value data collection with optimized utility,'' in
  \emph{USENIX Security Symposium}, 2020.

\bibitem{nie2019utility}
Y.~Nie, W.~Yang, L.~Huang, X.~Xie, Z.~Zhao, and S.~Wang, ``A utility-optimized
  framework for personalized private histogram estimation,'' \emph{IEEE TKDE},
  vol.~31, no.~4, pp. 655--669, 2019.

\bibitem{wang2017local}
S.~Wang, Y.~Nie, P.~Wang, H.~Xu, W.~Yang, and L.~Huang, ``Local private ordinal
  data distribution estimation,'' in \emph{IEEE INFOCOM}, 2017, pp. 1--9.

\bibitem{gu2019supporting}
X.~Gu, M.~Li, Y.~Cao, and L.~Xiong, ``Supporting both range queries and
  frequency estimation with local differential privacy,'' in \emph{IEEE CNS},
  2019, pp. 124--132.

\bibitem{gursoy2019secure}
M.~E. Gursoy, A.~Tamersoy, S.~Truex, W.~Wei, and L.~Liu, ``Secure and
  utility-aware data collection with condensed local differential privacy,''
  \emph{arXiv preprint:1905.06361}, 2019.

\bibitem{murakami2019utility}
T.~Murakami and Y.~Kawamoto, ``Utility-optimized local differential privacy
  mechanisms for distribution estimation,'' in \emph{USENIX Security
  Symposium}, 2019, pp. 1877--1894.

\bibitem{mcsherry2009privacy}
F.~D. McSherry, ``Privacy integrated queries: an extensible platform for
  privacy-preserving data analysis,'' in \emph{ACM SIGMOD}, 2009, pp. 19--30.

\bibitem{warner1965randomized}
S.~L. Warner, ``Randomized response: A survey technique for eliminating evasive
  answer bias,'' \emph{Journal of the American Statistical Association},
  vol.~60, no. 309, pp. 63--69, 1965.

\bibitem{jiang2018context}
B.~Jiang, M.~Li, and R.~Tandon, ``Context-aware data aggregation with localized
  information privacy,'' in \emph{IEEE CNS}, 2018, pp. 1--9.

\bibitem{cover2012elements}
T.~M. Cover and J.~A. Thomas, \emph{Elements of information theory}.\hskip 1em
  plus 0.5em minus 0.4em\relax John Wiley \& Sons, 2012.

\bibitem{huang2008optrr}
Z.~Huang and W.~Du, ``Optrr: Optimizing randomized response schemes for
  privacy-preserving data mining,'' in \emph{IEEE ICDE}, 2008, pp. 705--714.

\bibitem{itemset_data}
``Kosarak and retail datasets,'' \url{http://fimi.uantwerpen.be/data/}.

\bibitem{clothing}
``Clothing dataset,''
  \url{https://www.kaggle.com/rmisra/clothing-fit-dataset-for-size-recommendation}.

\end{thebibliography}

\end{document}